\documentclass[a4paper,11pt]{article}
\usepackage[utf8]{inputenc}

\usepackage[hidelinks]{hyperref}

\usepackage{framed}
\usepackage{mdframed}

\usepackage{soul}

\usepackage{fullpage}
\usepackage[T1]{fontenc}

\usepackage{amsfonts}

\usepackage[british]{babel}
\usepackage{verbatim}
\usepackage[T1]{fontenc}
\usepackage{lmodern}
\usepackage{lipsum}
\usepackage{booktabs}
\usepackage{caption}
\usepackage{cite}
\usepackage{soul,color}
\usepackage[toc,page]{appendix}
\usepackage{tikz,pgfplots}

\usepackage{ifpdf}

\usepackage{amsthm, amssymb}
\usepackage[tbtags]{amsmath}
\usepackage{bm}
\usepackage{mathtools}
\usepackage{amstext}
\usepackage{braket}
\usepackage{multirow}
\usepackage{hyperref}

\usepackage[normalem]{ulem}
\usepackage{xcolor}

\begin{document}
\vspace{5mm}
\vspace{0.5cm}

\def\be{\begin{eqnarray}}
\def\ee{\end{eqnarray}}

\def\ba{\begin{aligned}}
\def\ea{\end{aligned}}

\def\ls{\left[}
\def\rs{\right]}
\def\lc{\left\{}
\def\rc{\right\}}

\def\p{\partial}

\def\S{\Sigma}

\def\s{\sigma}

\def\O{\Omega}

\def\a{\alpha}
\def\b{\beta}
\def\g{\gamma}

\def\ad{{\dot \alpha}}
\def\bd{{\dot \beta}}
\def\gd{{\dot \gamma}}
\newcommand{\ft}[2]{{\textstyle\frac{#1}{#2}}}
\def\ib{{\overline \imath}}
\def\jb{{\overline \jmath}}
\def\Re{\mathop{\rm Re}\nolimits}
\def\Im{\mathop{\rm Im}\nolimits}
\def\trace{\mathop{\rm Tr}\nolimits}
\def\rmi{{ i}}
\def\N{\mathcal{N}}

\newcommand{\SU}{\mathop{\rm SU}}
\newcommand{\SO}{\mathop{\rm SO}}
\newcommand{\U}{\mathop{\rm {}U}}
\newcommand{\USp}{\mathop{\rm {}USp}}
\newcommand{\OSp}{\mathop{\rm {}OSp}}
\newcommand{\Symp}{\mathop{\rm {}Sp}}
\newcommand{\Sl}{\mathop{\rm {}S}\ell }
\newcommand{\Gl}{\mathop{\rm {}G}\ell }
\newcommand{\Spin}{\mathop{\rm {}Spin}}

\def\hc{c.c.}

\numberwithin{equation}{section}

\allowdisplaybreaks

\allowbreak

\setcounter{tocdepth}{2}


\begin{titlepage}
	\thispagestyle{empty}
	\begin{flushright}

		\hfill{LMU-ASC 05/22}
		
		\hfill{MPP-2022-20}

	\end{flushright}
\vspace{35pt}

	\begin{center}
	    { \Large\bf{
	   Large and Small Non-extremal Black Holes, 
	     }}
	     \vspace{10pt}
	     
	     { \Large\bf{
	Thermodynamic Dualities, and the Swampland
	     }}

		\vspace{50pt}

		{Niccol\`o~Cribiori$^{1}$, Markus~Dierigl$^{2}$, Alessandra~Gnecchi$^{1}$, Dieter~L\"ust$^{1,2}$ and Marco~Scalisi$^{1}$ }

		\vspace{25pt}

		{
			$^1${\it Max-Planck-Institut f\"ur Physik (Werner-Heisenberg-Institut),\\ F\"ohringer Ring 6, 80805, M\"unchen, Germany }

		\vspace{15pt}
            $^2${\it Arnold-Sommerfeld-Center for Theoretical Physics,\\ Ludwig-Maximilians-Universit\"at, 80333 M\"unchen, Germany }
 		}

		\vspace{40pt}

		{ABSTRACT}
	\end{center}

	\vspace{10pt}

In this paper we discuss black hole solutions parametrized by their entropy $\mathcal{S}$ and temperature $\mathcal{T}$ in gravitational effective theories. We are especially interested in the analysis of the boundary regions in the $\mathcal{T}-\mathcal{S}$ diagram, i.e.~large/small values of entropy and temperature, and their relation to Swampland constraints. To explore this correlation, we couple the gravitational theories to scalar fields and connect limits of thermodynamic quantities of black holes to scalar field excursions in the corresponding solutions. Whenever the scalar fields traverse an infinite field distance, the Swampland Distance Conjecture allows for a reformulation in terms of {\it entropy-} or {\it temperature-distance}. The effective theories with scalars  we investigate are Einstein-Maxwell-dilaton theory as well as $\mathcal{N} = 2$ supergravity in four dimensions. The relation of the latter to type II string theory compactified on Calabi-Yau 3-folds often allows for a direct identification of the corresponding light tower of states. These setups also point towards various dualities between asymptotic regions of the black hole solution. In the context of $\mathcal{N} = 2$ supergravity black holes, these {\it thermodynamic dualities} have an interpretation in terms of T- and S-dualities along the internal directions and their natural action on Kaluza-Klein and winding states.

\bigskip

\end{titlepage}


\def\thefootnote{\fnsymbol{footnote}}

\vskip 0.5cm

\vspace{0.5cm}

\def\thefootnote{\arabic{footnote}}
\setcounter{footnote}{0}

\baselineskip 5.6 mm


\tableofcontents


\newpage


\section{Introduction}

Swampland conjectures provide predictions for theories that are consistently coupled to gravity \cite{Vafa:2005ui,Ooguri:2006in,Palti:2019pca}. They can be seen as the imprint of the underlying quantum theory of gravity on the effective gravitational description at low energies. These predictions become trivial when gravity decouples, i.e., in the limit of vanishing Planck length $L_p =M_p^{-1}=\sqrt{8\pi \hbar G_N}\rightarrow 0$, that is either for small Newton constant $G_N$ or vanishing Planck constant $\hbar$.  Instead they tend to become more relevant in extremal regimes of the parameter space of the theory, for example when gauge couplings vanish, or scalar fields approach an infinite distance point in moduli space.

In this paper, we explore limiting regions of thermodynamic quantities of gravitational backgrounds and connect them to more familiar Swampland constraints. Most prominently, our investigation will focus on regions of large as well as small entropy $\mathcal{S}$ and temperature $\mathcal{T}$ of gravitational solutions within effective theories. Since we are interested in non-compact spacetimes, we restrict to situations where the temperature is associated to a horizon. We show that the Unruh temperature of Rindler spacetime does not lead to Swampland constraints. Black hole solutions, however, allow for a connection to Swampland constraints in certain limits of the thermodynamic properties captured by the Hawking temperature and entropy.

Our strategy is as follows. We describe black hole solutions in effective theories such as Einstein-Maxwell-dilaton theory and $\mathcal{N}=2$ supergravity that allow for limiting cases of vanishing and diverging $\mathcal{S}$ and $\mathcal{T}$. In situations where the effective theory allows for a string theory embedding we can also associate the limiting values of the thermodynamic order parameters with properties of the internal space, such as its volume. It turns out that in the string theory sub-class of these limiting solutions the moduli fields of the theory will traverse an infinite field distance from spatial infinity to the horizon.
In this way we are able to identify the tower of light string states in the vicinity of the black hole horizon, which is also suggested by the Swampland Distance Conjecture in the cases of infinite field distances.
Hence these limiting solutions need to be taken with care in view of the Swampland Distance Conjecture and are problematic within the effective description. This infinite field distance precisely happens in the limit of vanishing and diverging entropy of the black hole for arbitrary temperatures and thus offers a generalization of the investigations in \cite{Bonnefoy:2019nzv} for large entropies, see also \cite{DeBiasio:2020xkv,Luben:2020wix,Hamada:2021yxy}. We identify four generic limits in the thermodynamic quantities:

\begin{itemize}
\item{${\cal S}\rightarrow0$ and ${\cal T}\rightarrow \infty$: \, The moduli field distance diverges in this limit. The small horizon size and large temperature also suggests that quantum gravity effects become important. It was argued that this limit can be associated to elementary particles \cite{Holzhey:1991bx}.}
\item{${\cal S}\rightarrow\infty$ and ${\cal T}\rightarrow 0$: \, The moduli field distance diverges in this limit. This has been discussed in \cite{Bonnefoy:2019nzv} where the large number of black hole microstates was associated to a light tower of states.}
\item{${\cal S}\rightarrow0$ and ${\cal T}\rightarrow 0$: \, The moduli field distances diverges in this limit. Once more the vanishing horizon area suggests the importance of quantum gravity corrections to avoid the violation of entropy bounds, see \cite{Hamada:2021yxy}. As we will show this behavior is general for arbitrary finite temperatures $\mathcal{T}$ and the associated non-extremal black holes.}
\item{${\cal T}\rightarrow 0$ and $\mathcal{S}$ finite: \, The moduli field distance is finite in this limit. Particularly this regime includes the charged extremal black holes, which can be described within the effective theory. Note, however, that in the non-supersymmetric case it was argued that there still might be large quantum corrections for the thermodynamic quantities \cite{Preskill:1991tb, Maldacena:1998uz, Page:2000dk, Heydeman:2020hhw}.}
\end{itemize}

These results confirm the importance of quantum gravity corrections in these situations and thus provides an interesting correlation between black hole thermodynamics and Swampland Constraints, suggesting a notion of {\it entropy} and {\it temperature-distance}\footnote{See also \cite{Agrawal:2020xek}, which suggests that the limit of high temperature should be at infinite distance.}. Note that for this observation it is crucial that the string theory black hole geometries are described by warped products, as opposed to fibrations, and therefore we do expect the modes arising from compactification to be stable\footnote{For a non-trivial fibration it is often the case that winding modes can unwrap and the Kaluza-Klein momentum is not preserved. In such situations we do not expect a stable tower of states, see e.g. \cite{Draper:2019utz, Lanza:2021udy}}.

We also investigate the black hole solutions above in terms of a distance in the space of metrics (see for example \cite{PhysRev.160.1113}), which diverges in all the limiting cases. An explanation for this can be the formation of an infinite throat in the black hole geometry (see also  \cite{Li:2021gbg,Li:2021utg}  for a related discussion and \cite{Elander:2020rgv} for an holographic application). We therefore conclude that the metric distance as a Swampland parameter in these non-compact backgrounds needs to be taken with care.  

Moreover, the black hole formulae suggest that some of the limits might be connected by a form of {\it thermodynamic dualities} which for example act as $\mathcal{T} \rightarrow 1/\mathcal{T}$ or $\mathcal{S} \rightarrow 1/\mathcal{S}$. We regard this as a formal result, which is expected not to hold in its original form once corrections are taken into account (e.g. in the limit of low entropy, one expects the contribution of higher derivative terms). Nevertheless, in the string embedding, we will be able to relate these to a combination of T- and S-dualities affecting the internal volume as well as the coupling constants. From the point of view of the Swampland program, the emergence of these dualities provides additional testament that duality itself should be a general principle of quantum gravity. While this is one of main lessons of string theory, our investigation suggests a novel manifestation of this in terms of the thermodynamic properties within a bottom-up description. 

The paper is organized as follows. In Section \ref{sec:BHgen} we start by discussing temperature limits in gravity, for Rindler space and also for the Schwarzschild black hole. Then we review relevant aspects of the Reissner-Nordstrom black hole as well as the notion of distance in the space of black hole geometries. In Section \ref{sec:EMDblackholes} we investigate in detail non-extremal dyonic black holes, coupled to a dilaton field. In this context we study the $\mathcal{T}-\mathcal{S}$ phase diagram of the charged black holes and the various extremal limits. In Section \ref{SG-blackholes} we turn to the ${\cal N}=2$ supergravity realization of general four-dimensional, non-extremal black holes together with their behavior under a variation of thermodynamic quantities. Focussing on the $STU$ model we can explicitly identify the tower of states. We further show the supergravity embedding of the Einstein-Maxwell-dilation black hole into ${\cal N}=2$ supergravity. In Section \ref{sec:concl} we present our conclusions. Some generalities about $\mathcal{N} = 2$ supergravity  are collected in the Appendix.
                                
\section{Distance, temperature and entropy in gravity}
\label{sec:BHgen}

In the context of the Swampland Program \cite{Vafa:2005ui,Ooguri:2006in,Palti:2019pca}, large variations of parameters in an effective theory become problematic when consistently coupling the theory to gravity. In particular, the Swampland Distance Conjecture states that a large variation of field values for a scalar field\footnote{The correct quantity is the geodesic distance in the scalar moduli space.}  $\phi$ is accompanied by a light tower of an infinite number of states. These light states in turn invalidate the effective description by lowering the quantum gravity cut-off. Specifically, the masses of the states in the tower decrease exponentially in the field distance $\Delta (\phi)$
\begin{align}
{m L_p} \sim e^{-\lambda \Delta(\phi)} \,,
\label{gendisca}
\end{align}
where $\lambda \sim \mathcal{O}(1)$ is a positive parameter. For this to be a meaningful Swampland constraint the tower needs to disappear when decoupling gravity, i.e., for $L_p \rightarrow 0$. This is realized if the masses of the states in the tower remain positive when measured in Planck units, i.e., 
\begin{align}
m L_p = \frac{m}{M_p} > 0 \quad \text{for} \quad L_p \rightarrow 0 \,.
\label{eq:qgconst}
\end{align}
If this is not the case, the related tower cannot be identified with the states postulated by the SDC.

In this section, we explore whether also the variation of thermodynamic quantities such as temperature $\mathcal{T}$ and entropy $\mathcal{S}$ can lead to Swampland constraints, similar to the SDC. For that we first investigate flat space at finite temperature and then turn to the analysis of black hole geometries.

\subsection{Temperature in flat space}

One natural situation where the SDC becomes relevant are field theories derived by compactification. When the internal space becomes large the corresponding momentum modes, the Kaluza-Klein (KK) states, become light constituting the predicted tower. In the opposite limit, when the internal space becomes small, one expects dual winding modes to appear. This seems to have  direct implications for field theories at finite temperature. However, in the following we will see that the temporal circle in flat space is not associated to quantum gravity constraints imposed by Swampland conjectures.

To describe the finite temperature regime one performs a Wick rotation of the time direction, $t \rightarrow i \tau$, and compactifies the Euclidean time on a circle. The radius of this circle $R_{\tau}$ is related to the temperature $\mathcal{T}$ of the system\footnote{We set the Boltzmann constant to one.}
\begin{align}
\mathcal{T} = \frac{\hbar}{\beta} = \frac{\hbar}{2 \pi R_{\tau}} \,.
\end{align}
In analogy to the KK states in a circle compactification, we can associate a tower of states with frequencies
\begin{align}
\omega_n = \frac{2 \pi n}{\beta} = \frac{2 \pi n}{\hbar} \mathcal{T} \,,
\end{align}
to the time-like circle. These are the so-called {\it Matsubara modes}. One then might be tempted to predict a tower of states with mass scale determined by
\begin{align}
m = \omega_1 = \frac{2 \pi}{\hbar} \mathcal{T} \,.
\label{eq:Matfreq}
\end{align}
This would follow from a distance associated to a variation of the temperature given by
\begin{align}
\Delta_{\mathcal{T}} \sim \bigg| \frac{1}{\lambda} \text{log} \Big( \frac{2 \pi L_p}{\hbar} \, \mathcal{T} \Big) \bigg| \,.
\end{align}
The SDC then demands the appearance of a light tower for $\Delta_{\mathcal{T}} \rightarrow \infty$. One realization of this limit is given by $\mathcal{T} \rightarrow 0$, with the Matsubara modes above becoming light. Another realization can be obtained by $\mathcal{T} \rightarrow \infty$, which would demand a dual tower of thermal winding modes, see \cite{Atick:1988si,Kounnas:1989dk,Angelantonj:2008fz}. Therefore, these considerations show that the temperature might indeed be an interesting parameter to explore from a quantum gravity perspective.

However, a gravitational system at finite temperature is highly problematic \cite{Atick:1988si}. The finite temperature induces a finite energy density which, in a large enough region, leads to gravitational (Jeans) instabilities. Since we are still interested in non-compact backgrounds, we explore other gravitational systems that allow for a well-defined notion of temperature. In these setups the temperature appears at a horizon. For these backgrounds, we investigate the appearance of light states, similar to the Matsubara modes above, in the extremal temperature regime $\mathcal{T} \rightarrow 0$ and $\mathcal{T} \rightarrow \infty$. Importantly, all meaningful towers need to satisfy \eqref{eq:qgconst}.

As a warm-up example we can study the temperature observed by an accelerated observer, i.e., the Unruh effect in Rindler space. Rindler space can be obtained from Minkowski spacetime with coordinates $(x_1, x_2, x_3, x_4)$ by the following identification
\begin{align}
x_0=\rho\sinh (a\eta)\, ,\qquad x_1=\rho\cosh(a \eta) \,.
\end{align}
In these hyperbolic coordinates the four-dimensional metric is given by
\begin{align}
ds^2=(-a^2\rho^2d\eta^2+d\rho^2)+(dx_2)^2+(dx_3)^2 \,,
\label{Rindler}
\end{align}
where $a$ is the acceleration parameter of the observer. Due to the Unruh effect, the accelerated observer experiences a thermal radiation, and the corresponding Unruh temperature can be read off from the Rindler metric by switching to Euclidean time, i.e., by replacing $\eta= i\tau$. To avoid a conical singularity at the origin, the Euclidean time $\tau$ must have periodicity $\tau\sim\tau+\beta_U$ and the Unruh temperature is given by
\begin{align}
{\cal T}_U =     {\hbar \over\beta_U}={\hbar a\over 2\pi} \,.
\end{align}
In analogy to the Matsubara frequency \eqref{eq:Matfreq}, we define the Unruh frequency
\begin{align}
\omega_U = \frac{2 \pi}{\hbar} \, \mathcal{T}_U = a \,.
\end{align}
We see that $\omega_U$ and therefore the associated mass scale does not depend on $L_p$. Consequently, the decoupling condition \eqref{eq:qgconst} is not satisfied for finite $a$ and one does not obtain an SDC tower by varying $\mathcal{T}_U$. This is indeed expected, since Rindler space describes part of flat space. However, it shows that the condition \eqref{eq:qgconst} is crucial in order to draw conclusions concerning Swampland constraints associated to the temperature of a system.

Therefore, we turn to different gravitational backgrounds that exhibit a temperature and have a more direct relation to quantum gravity, namely black hole metrics. We further include a second thermodynamic quantity, the entropy, which allows the study of a two-parameter family of gravitational solutions.

\subsection{Schwarzschild black holes and Hawking temperature}

We start by considering the simplest black hole solution, namely the Schwarzschild black hole, whose metric is given by
\begin{equation}
\text ds^2 = -f(r)\text dt^2+ f(r)^{-1} \text dr^2+r^2\text d\Omega_{2}^2\;,\qquad f(r) = 1-\frac{2M G_N}{r} \,,
\label{staticbh}
\end{equation}
with mass $M$ and horizon size $r_S=2MG_N$. The associated Bekenstein-Hawking entropy reads
\begin{equation}
\label{entropySchw}\boxed{
{\cal S}= {8 \pi^2 r_S^2 \over  L_p^2} = {32 \pi^2 G_N^2 M^2 \over  L_p^2} = \frac{L_p^2 M^2}{2 \hbar^2} \,,}
\end{equation}
and the Hawking temperature is 
\begin{equation}\boxed{
{\cal T} = { \hbar\kappa\over 2\pi}={\hbar \over 8\pi G_N M} = {\hbar^2\over L_p^2 M} \,,}
\label{HawkingT}
\end{equation}
where $\kappa = (2 r_S)^{-1}$ is the surface gravity.

For the Schwarzschild black hole, the entropy and the Hawking temperature are not independent quantities but are related as (setting $\hbar=L_p=1$)
\begin{equation}
{\cal S}{\cal T}^2 = \frac{1}{2} \,.
\label{STrel}
\end{equation}
Anticipating some later results we further note that the transformation 
\begin{equation}
\sqrt{\cal S}\,\longleftrightarrow\,{\cal T}
\label{STex}
\end{equation}
 exchanges small with large black holes, namely acts on the mass as $M\longleftrightarrow{1\over M}$, while leaving eq.~\eqref{STrel} invariant. Using the relation (\ref{STrel}) we can express the transformation eq.~\eqref{STex} (up to a constant factor) equivalently as
\begin{equation}
{\cal T}\,\longleftrightarrow\,{1\over {\cal T}} \quad{\rm and}\quad {\cal S}\,\longleftrightarrow\,{1\over {\cal S}}\, .
\label{STdual}
\end{equation}
In section (\ref{thlimits}) we will discuss how these transformations act on the internal KK and winding spectrum in supergravity and in string compactifications. 

It is instructive to recall how the Hawking temperature can be derived from the near horizon geometry. For this purpose, we introduce the  coordinate ${x^2\over 8G_NM}=r-r_S$ and for small $x^2$ we obtain the metric in the near horizon limit and for Euclidean time $\tau=it$,
  \begin{equation}
\text ds^2=(\kappa x)^2\text d\tau^2+ dx^2+(2\kappa)^{-2}\text d\Omega_2^2\,. 
\label{nearHorizonMetricSchwarzschild}
 \end{equation}
The two-dimensional part of this metric for $x$ and $\tau$ is the Rindler space metric \eqref{Rindler} and we can immediately read off the Hawking temperature ${\cal T}$ in agreement with \eqref{HawkingT}. Notice that the derivation is completely analogous to that of the Unruh temperature, i.e., we require the absence of a conical singularity at the origin. However, while in the case of the Unruh temperature the acceleration $a$ is a free parameter, in the case of the Hawking temperature the surface gravity $\kappa$ fixes also the size of the $S^2$ part of the space. As such, $\kappa$ is not a free parameter, but it is determined by the geometry of the full four-dimensional space. This has important consequences and in fact amounts to saying that, for the case of Schwarzschild black hole, entropy and temperature are interdependent quantities.

The two potentially interesting temperature regimes $\mathcal{T} \rightarrow \{ 0, \infty\}$ can be rephrased in terms of the mass $M$ of the black hole:\footnote{In addition there are also two kinds of classical limits:
\vskip0.2cm\noindent
Semiclassical limit: $M\rightarrow\infty$ and $G_N\rightarrow 0$ while keeping $\hbar$ and $r_S$ finite. One has that $L_p \rightarrow 0$. In this limit, ${\cal T}$ stays finite and the Hawking radiation is exactly thermal. 
\vskip0.2cm\noindent
Classical limit:  $\hbar \rightarrow 0$ while keeping $M$ and $G_N$ finite, and hence also $r_S$ finite. Again, one has that $L_p \rightarrow 0$. In this limit, the Hawking temperature goes to zero, ${\cal T}\rightarrow 0$, and therefore there is no Hawking radiation.
 \vskip0.2cm\noindent
For these two cases the Planck length goes to zero, since either gravity decouples  or because quantum effects vanish. Therefore, for these two cases Swampland arguments should not be applicable and hence we do not generically expect a massless tower of states.} 

\vskip0.2cm
(I) Small black hole limit,  $\hbar$ finite, $G_N$ finite, $M\rightarrow 0$: \,  One has $r_S\rightarrow 0$ ($\kappa\to\infty$). In this limit, the Hawking temperature goes to infinity, ${\cal T}\rightarrow \infty$, and at the same time the entropy becomes very small, i.e. ${\cal S}\rightarrow 0$. For $M=0$ one is dealing with flat Minkowski spacetime.  

\vskip0.2cm
(II) Large black hole limit, $\hbar$ finite, $G_N$ finite, $M\rightarrow \infty$: \,  One has $r_S\rightarrow \infty$ ($\kappa\to0$). In this limit, the Hawking temperature goes to zero, ${\cal T}\rightarrow 0$, and therefore there is no Hawking radiation. The entropy becomes very large, ${\cal S}\rightarrow\infty$, and the near-horizon geometry approaches Rindler space with zero temperature, which is again flat Minkowski spacetime.

\vskip0.2cm
\noindent
Since in both of these limits $L_p$ remains finite, we expect quantum gravity effects to be important and Swampland constraints to be applicable.

Let us once more use the analogy to finite temperature systems giving rise to Matsubara modes, by defining the Hawking frequency via the Hawking temperature
\begin{align}
\omega_H={2\pi\over \hbar }{\cal T}=\kappa={2 \pi \hbar\over L_p^2M} \,.
\label{eq:Hfreq}
\end{align}
In the limit $\mathcal{T} \rightarrow 0$, which is limit (II) above, also the Hawking frequency vanishes. However, as opposed to the Unruh temperature the consistency condition \eqref{eq:qgconst} is satisfied and the mass scale decouples for $L_p \rightarrow 0$. This indicates that the temperature of a black hole solution might be a viable Swampland parameter. Utilizing \eqref{STrel} we can further determine the relevant quantities in terms of the entropy $\mathcal{S} = \frac{L_p^2 M^2}{2}$ of the Schwarzschild black hole
\begin{align}
\omega_H=\frac{\sqrt{2} \pi}{L_p \sqrt{\cal S}} \,.
\end{align}
This demonstrates that the vanishing of the Hawking frequency corresponds to the limit of infinite entropy $\mathcal{S} \rightarrow \infty$ which was also discussed in \cite{Bonnefoy:2019nzv}. Assuming that a tower of the mass scale \eqref{eq:Hfreq} indeed appears, this can be used to define a distance for Schwarzschild black hole geometries given by
\begin{align}
\Delta_{BH}=\left|{1\over \lambda}\log\biggl({2\pi L_p\over \hbar}{\cal T}\biggr)\right|=\left|{1\over \lambda}\log \biggl(\frac{\sqrt{2} \pi}{  \sqrt{\cal S}}\biggr)\right| \,.
\label{BHDIST}
\end{align}
In \cite{Bonnefoy:2019nzv} it was argued that at infinite entropy, i.e., for $\Delta_{BH} \rightarrow \infty$ an infinite number of black hole microstates becomes degenerate. Interpreting $\omega_H$ as a natural mass gap between those, therefore also leads to an interpretation in terms of a light tower of states. 
However a priori there is nothing quantum-mechanical or gravitational about this: the black hole becomes large, so its characteristic frequency goes to zero. 
Therefore, as already mentioned, we like to consider black hole solutions, for which ${\cal S}$ and ${\cal T}$ are related to moduli parameters and to the physical towers of scalar fields within an underlying string theory
realization.

This infinite entropy limit can also be discussed from the point of view of the near horizon geometry, which is ${\cal R}^{1,1}\times S^2$, where ${\cal R}^{1,1}$ is the two-dimensional Rindler space and $S^2$ is the 2-sphere. In the limit ${\cal S}\rightarrow \infty$ (${\cal T}\rightarrow 0$), the volume of the $S^2$ becomes infinite and the corresponding Kaluza-Klein modes becomes massless. This could be a signal of a  breakdown  of the two-dimensional effective field theory on ${\cal R}^{1,1}$. However, it is unclear if one should apply Swampland considerations in two (and three) dimensions, and it would be safer to derive Swampland arguments signaling a breakdown of the entire four-dimensional effective field theory. 

The black hole distance $\Delta_{BH}$ in \eqref{BHDIST} also diverges in the \emph{dual} limit of small entropy and large temperature ${\cal S}\rightarrow 0$, ${\cal T}\rightarrow \infty$, corresponding to case (I). We could thus wonder if there are certain modes which become massless in this limit. We will come back to this question in the context of charged dilatonic and supergravity black holes.

\subsection{Reissner--Nordstrom black hole, temperature, and entropy}

Motivated by the previous discussion, we extend our consideration in this section to black holes with two parameters, mass and charge, i.e., Reissner-Nordstrom black holes. For a black hole with non-vanishing charge, we can introduce the concept of extremality. A black hole is extremal if it has the lowest possible mass for a fixed charge, while avoiding naked singularities. The Reissner--Nordstrom (RN) solution is a charged black hole with metric
\begin{equation}
\label{RNbh}
\text ds^2 = -f(r) \text dt^2 + f(r)^{-1} \text dr^2 + r^2  \text d \Omega_2^2, \qquad f(r) = 1 - \frac{2M G_N}{r} + \frac{Q^2 G_N}{r^2} \,,
\end{equation}
where $M$ is the mass and $Q$ the (electric) charge.  This black hole has two horizons located at the zeroes of the function $f(r)$,
\begin{equation}
f(r_\pm) = 0, \qquad r_\pm =  G_N( M \pm c) \,, 
\end{equation}
where we introduced the extremality parameter $c$ such that
\begin{equation}
G_N\,c  = \sqrt{G_N^2 M^2- G_NQ^2} \geq0 \,, \qquad M\geq c \geq 0 \,.
\end{equation}
In order to avoid naked singularities, we have to require $M^2\geq Q^2$. The entropy of the black hole is given by
\begin{equation}
\label{SRN}\boxed{
\mathcal{S} = \frac{8 \pi^2 r_+^2}{L_p^2} \,.}
\end{equation}
It is important to notice that the regimes $c>0$ and $c=0$ are qualitatively different, so we discuss them separately below.

For $c >0$, we can introduce the coordinate $\frac{G_N\, c \, x^2}{2 r_+^2} = r-r_+$. For small $x^2$ we obtain the metric in the near horizon limit and with Euclidean time $\tau = i t$, 
\begin{equation}
\label{RNnh}
\text d s^2 =  (\kappa x)^2 \text d \tau^2 + \text d x^2 + (r_+)^2 \text d \Omega_2^2 \,,
\end{equation}
where the surface gravity is $\kappa = \frac{G_N\, c}{r_+^2}$.
By imposing the absence of conical singularities, we find the temperature
\begin{equation}
\label{TRN}\boxed{
{\cal T}= \frac{\hbar\kappa}{2 \pi}  = \frac{\hbar c}{2 \pi G_N (M+c)^2} \,.}
\end{equation}
In terms of the Hawking temperature ${\cal T}$ and the entropy ${\cal S}$, the extremality parameter can be expressed as
\begin{equation}
\label{c=2ST}\boxed{
c = 2 \mathcal{S} {\cal T} \,,}
\end{equation}
and the metric \eqref{RNnh} can be rewritten as
\begin{equation}
\label{RNnh1}
\text d s^2 =  \Big({4\pi^2\over \hbar^2}\Big)({\cal T} x)^2 \text d \tau^2 + \text d x^2 + \Big({L_p^2\over 8 \pi^2}\Big){\cal S} \,\text d \Omega_2^2 \,.
\end{equation}
Therefore, the radius of the Euclidean time-circle is given in terms of ${\cal T}$, whereas the area of the $S^2$ is determined by the entropy ${\cal S}$. This has to be contrasted with the Schwarzschild black hole, where both these quantities were governed by the same (unique) parameter $M$ of the solution. The Hawking frequency of the RN black hole is given by
\begin{equation}
\label{matsRN}
\omega_H = \frac{2\pi}{\hbar} {\cal T} = \kappa = \frac{G_N\, c}{r_+^2} \, ,
\end{equation}
and again the condition \eqref{eq:qgconst} is satisfied. Therefore, the variation of the temperature of a RN black hole might lead to interesting constraints.

For the extremal case, $c=0$, the coordinate $x^2$ introduced above becomes ill-defined. Thus, we cannot take the limit $c \rightarrow 0$ in the near horizon metric \eqref{RNnh}. The correct strategy is to first take the extremal limit and then the near horizon limit. The RN metric \eqref{RNbh} then reduces to
\begin{equation}
\text d s^2 = - \Big(\frac{r-r_h}{r}\Big)^2 \text d t^2 + \Big(\frac{r-r_h}{r}\Big)^{-2} \text dr^2 + r^2 \text d \Omega_2^2,
\end{equation}
where $r_h = G_N M = G_N^{1/2} Q$ is the horizon radius (we will set $M_p=1$ from now on). The near horizon coordinate is now $x = r-r_h$ and for small $x$ we get the metric 
\begin{equation}
\label{RNextnh}
\text ds^2 = - \frac{x^2}{r_h^2} \text d t^2 + \frac{r_h^2}{x^2} \text d x^2 + r_h^2 \text d \Omega_2^2.
\end{equation}
We recognize the $AdS_2$ metric along the $(t,x)$ directions with radius $r_h$. This metric is clearly regular everywhere, so there is no temperature. Equivalently, taking ${\cal T} = \frac{\hbar \kappa}{2 \pi}$ as a definition of Hawking temperature, we can calculate the surface gravity for the extremal RN black hole and find that $\kappa=0$, implying ${\cal T}=0$.\footnote{We recall that for Schwarzschild and RN black holes the surface gravity is given by $\kappa = \frac12 f^\prime(r_+)$}

\begin{figure}
\begin{center}
\begin{tikzpicture}[scale=1.25]
  \draw[->, line width=.5mm] (0, 0) -- (4.2, 0) node[below] {${\cal T}$};
  \draw[->,line width=.5mm] (0, 0) -- (0,4.33) node[left] {${\cal S}$};
\draw[scale=1.5,domain=0.6:2.7,smooth, variable=\x,blue, thick,line width=.6mm] plot ({\x},{1/((\x)*(\x))});
\fill [blue!25, domain=0.1:4.2, variable=\x]
  (0.02, 4.17)
  -- plot[scale=1.5,domain=0.6:2.7] ({\x},{1/((\x)*(\x))})
  -- (0.02, 0.02);
  \fill [blue!25] (0.02,0.02) rectangle (4.05,0.2); 
  \node () at (3,3) {excluded region};
  \draw[scale=1.5,domain=1.04:2.7,smooth, variable=\x,red,line width=.6mm] plot ({\x},{1/(\x)});
  \draw[scale=1.5,domain=0.36:1.1,smooth, variable=\x,red, dashed,line width=.6mm] plot ({\x},{1/(\x)});
  \draw[dashed] (1.55,1.4)--(1.55,0);
  \node () at (1.55,-.3) {${\cal T}_0$};
\end{tikzpicture}
\end{center}
\caption{Population of the ${\cal T}-\mathcal{S}$ plane by RN black holes. The boundary ${\cal S}=(2 {\cal T}^2)^{-1}$ between the allowed region in blue (with $Q^2>0$)
 and the excluded region (where
$Q^2<0$) is given by uncharged ($Q=0$) Schwarzschild black holes (blue line). The red line is the line of constant extremality parameter $c=2 \mathcal{S} {\cal T}$. For ${\cal T}>{\cal T}_0={1\over c}$, one cannot keep $c$ constant while staying in the allowed region.}
\label{fig:BHpopulation}
\end{figure}
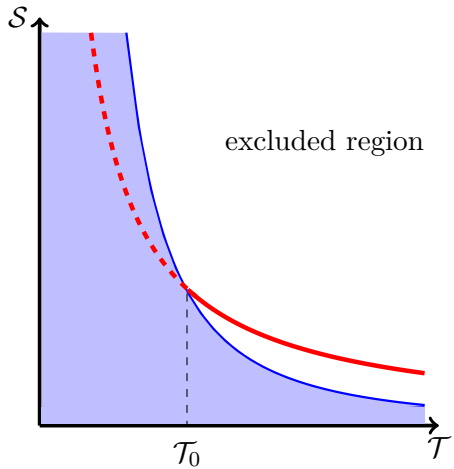

For the purposes of our analysis, it is convenient to choose a basis for the parameter space of RN solutions in which the independent coordinates are ${\cal T}$ and $\mathcal{S}$. We can then study the limit of small or large temperature or entropy. Using, \eqref{SRN}, \eqref{TRN} and \eqref{c=2ST}, we can rewrite the full RN metric in terms of ${\cal T}$ and $\mathcal{S}$ as
\begin{equation}
\text ds^2 = - \frac{(r-r_+)(r-r_-)}{r^2} dt^2+\left(\frac{(r-r_+)(r-r_-)}{r^2}\right)^{-1} dr^2 + r^2 d\Omega_2^2,
\end{equation}
where
\begin{equation}
r_+ =  \sqrt{\frac{\mathcal{S}}{8 \pi^2}},\qquad r_- = \sqrt{\frac{\mathcal{S}}{8 \pi^2}} -  \frac{1}{2 \pi} \mathcal{S}{\cal T}.
\end{equation}
In the small temperature limit, ${\cal T} = 0$, the expressions simplify. Clearly, this basis illustrates that temperature and entropy of a RN black hole are different order parameters. In a ${\cal T}-\mathcal{S}$ diagram, charged RN black holes populate the region bounded by uncharged Schwarzschild black holes, as shown in Figure \ref{fig:BHpopulation}.

While the condition \eqref{eq:qgconst} is satisfied for the Hawking temperature, it is not clear whether a distance defined by \eqref{BHDIST} and its related version for the RN black hole is appropriate for an interpretation in terms of the SDC. The reason for this is that we employed an analogy to the Matsubara and therefore KK momentum states that appear if the underlying geometry contains a circle. However, in the black hole geometries there is in general no such a circle and one has to be more careful. The distance \eqref{BHDIST} would for example demand that whenever a black hole becomes extremal there should be an associated light tower of states. This seems to be unlikely, since for supersymmetric black holes we even control the microscopic details and small deformations away form extremality seem to be well-behaved, see e.g. \cite{Saraikin:2007jc}. Moreover, it was argued that the semi-classical thermodynamic quantities in non-supersymmetric setups are potentially subject to large quantum corrections in the extremal limit \cite{Preskill:1991tb, Maldacena:1998uz, Page:2000dk, Heydeman:2020hhw}.

Our strategy is therefore to relate the black hole limits of large/small temperature and entropy to field excursions of scalar fields, which are under better control and have a direct relation to the SDC. In order to do so, we need to extend our theories by coupling to scalar fields. This is why we turn to Einstein-Maxwell-dilaton theory in Section \ref{sec:EMDblackholes} and to the $STU$ model in $\mathcal{N} = 2$ supergravity in Section \ref{SG-blackholes}. 

Before we start our discussion with scalar fields, we want to analyze the two-parameter black hole solutions above from the perspective of a distance in the space of metrics.

\subsection{Distance in the space of metrics}

One possible definition of a distance between distinct solutions to the gravitational equations of motion is given by the metric distance as discussed in \cite{Bonnefoy:2019nzv}. In practice, we can consider these distances directly in the near horizon geometries. We have seen that starting from the RN black hole \eqref{RNbh} and performing first the near horizon limit, we get the geometry \eqref{RNnh}, which is locally ${\cal R}^{1,1} \times S^2$, with non-vanishing Hawking temperature. The extremal limit is then ill-defined, since the near horizon coordinate becomes singular for $c=0$. On the other hand, performing first the extremal limit and then the near horizon limit, we arrive at the metric \eqref{RNextnh}, which is locally $AdS_2 \times S^2$ and which has vanishing temperature. 
The question is then if the spaces  ${\cal R}^{1,1} \times S^2$ and $AdS_2 \times S^2$ could be at infinite distance to one another, with the Hawking modes \eqref{matsRN} being the tower predicted by the SDC.

Given a family of metrics with a set of parameters which we collectively denote by $\alpha$, the distance in the parameter space is defined via the formula
\begin{equation}
\label{distmetric}
\Delta_g(\alpha)  \sim \left|\int_{\alpha_i}^{\alpha_f} \text d \alpha \left(\frac{1}{\text{Vol}} \int d^4 x \sqrt{g} \, g^{\mu\nu}g^{\rho\sigma}\frac{\partial g_{\mu\rho}}{\partial \alpha} \frac{\partial g_{\nu\sigma}}{\partial \alpha}\right)^{\frac 12}\right|,
\end{equation}
where $\text{Vol} = \int \text d^4x \sqrt{g}$ is the volume of spacetime. We can now use this formula for different choices of the parameters characterizing the solution.

The simplest choice is to consider the metric \eqref{RNnh} as a function of the extremality parameter $c$. We can compute the integral above and find
\begin{equation}
\Delta_{RN,nh} (c) \sim |\log c|\, \qquad \text{for} \quad c \to 0 \,. 
\end{equation}
Therefore, taking the extremal limit in the near horizon RN metric \eqref{RNnh} corresponds to going to infinite distance in the space of metrics. One possible explanation for this divergence in the space of metrics is the formation of an infinitely long throat. Indeed, if one calculates the distance to the outer horizon from a reference radius $r_*$ outside the black hole one finds (along a slice of constant time)
\begin{align}
\ell = \int_{r_+}^{r_*}  \sqrt{g_{rr}} \,dr =\int_{r_+}^{r_*} dr \sqrt{\frac{r^2}{(r - r_+) (r - r_-)}} \, ,
\end{align}
which is finite for non-extremal black holes, but it diverges logarithmically in the extremal case. This can be seen explicitly in the shifted radial coordinate $y = r - r_+$, giving
\begin{align}
\ell = \int_{0}^{r_* - r_+} dy \sqrt{ \frac{(y + r_+)^2}{y \big(y + \tfrac{c}{8 \pi} \big)}} \,,
\end{align}
leading to a logarithmic contribution $\sim|\log c \, |$ close to the outer horizon, just as for the metric distance. Therefore, a conceivable physical interpretation is that the infinite metric distance is induced by the stretching of the throat region when sending $c \rightarrow 0$.

We can also consider ${\cal S}$ and ${\cal T}$ as independent parameters and repeat a similar analysis. The metric distance with two parameters $\alpha_i=(\alpha_1,\alpha_2)$ along a path $\alpha(\alpha_i)$ is 
\begin{align}
\label{distance}
\Delta(g) \sim \left|\int_{\alpha_i}^{\alpha_f}d\alpha \left(
\frac{1}{\text{Vol}}\int d^4 x \sqrt{g} \, g^{\mu\nu}g^{\rho\sigma}\partial_i g_{\mu\rho}\partial_j g_{\nu\sigma}\right)^{1/2}\right|
\end{align}
It is now convenient to employ the Kruskal extension of the black hole metric at the outer horizon
\begin{align}
\label{Kruskal-RN-outer-hor}
ds^2_{Kruskal}&=- \frac{1}{2\pi^2 {\cal T}^2} dU dV + \frac{\cal S}{8 \pi^2} d\Omega_{2}^2  \ .
\end{align}
Using these coordinates and for $\alpha_1 = {\cal T}$, $\alpha_2=\mathcal{S}$, we find
\begin{equation}
\Phi_{ij} [g_{\mu\nu}, \alpha^k]\equiv \frac{1}{\rm Vol}\int d^4x \sqrt{g} \, g^{\mu\nu} \partial_ig^{\rho\sigma}\partial_{j}g_{\mu\rho} = 
\left(\begin{array}{cc}
\frac{8}{{\cal T}^2}&0 \\ 
0 & \frac{2}{{\cal S}^2}
\end{array}
\right)
\end{equation}  
and thus for a path parametrized by $\tau \in [0,1]$
\begin{align}\label{metricdistanceRN}
\Delta(g)&\sim  \int_0^1 d\tau \sqrt{\Phi_{ij} \frac{d \alpha^i}{d\tau} \frac{d \alpha^j}{d \tau}}  = \int_0^1 d \tau \sqrt{8 \Big(\frac{d \, \text{log} \mathcal{T}}{d\tau}\Big)^2 + 2 \Big( \frac{d \, \text{log} \mathcal{S}}{d \tau} \Big)^2 } \,,
\end{align}
for the metric in the two parameter space of the RN black hole. This distance\footnote{The specific distance depends on the path chosen.} diverges in the limits $\mathcal{S}, \mathcal{T} \rightarrow \{ 0, \infty\}$.

The metric distance \eqref{metricdistanceRN} in fact diverges in any of the five limits discussed in the introduction of this paper. Nevertheless the identification of the corresponding tower of states represents in general a challenge. While the stretching of a throat at first might have some similarity to a decompactification process, which leads to a tower of light KK modes, the direct correlation to the SDC is more subtle, see also \cite{Li:2021utg}. For the special limit $\mathcal{S} \rightarrow \infty$ there are other checks that suggest the appearance of light modes, i.e. the high ground state degeneracy. The same logic, however, cannot be applied for finite $\mathcal{S}$. Therefore, we do not conclude that extremal $\mathcal{T} = 0$ black holes are at infinite distance for finite entropy, but rather try to explore these limits in systems with scalar fields. This points towards limitations of the metric distance as a Swampland parameter, resonating with the additional caveat that \eqref{distmetric} for the metric distance is not diffeomorphism invariant\footnote{A procedure to deal with this technical obstruction has been proposed in \cite{Bonnefoy:2019nzv}, to which we refer the reader.}.

\section{Dyonic black holes in Einstein-Maxwell-dilaton theory}
\label{sec:EMDblackholes}

In this section, we investigate black holes with electric and/or magnetic charges in Einstein-Maxwell-dilaton (EMd) theory. We start with a review of these solutions, following \cite{Ivashchuk:1999jd, Abishev:2015pqa, Loges:2019jzs} (see also \cite{Gibbons:1987ps, Garfinkle:1990qj, Cheng:1993wp, Heidenreich:2015nta}). The action is given by
\begin{align}
S_{\text{EMd}} = \int d^4 x \sqrt{- g} \Big( \tfrac{1}{2} M^2_p \big( R - \partial_{\mu} \phi \partial^{\mu} \phi \big) - \tfrac{1}{4} e^{-2 \lambda \phi} F_{\mu \nu} F^{\mu \nu} \Big) \,,
\end{align}
with dilaton field $\phi$ and its coupling to the U$(1)$ gauge field determined by $\lambda$. In particular, the previously considered RN black hole corresponds to $\lambda=0$.
It is convenient to introduce a rescaled dilaton coupling,
\begin{align}
h = \frac{2}{1+ 2 \lambda^2} \leq 2 \,,
\end{align}
with respect to which the static, spherically symmetric black holes solutions can be given as
\begin{equation}
\begin{split}
ds^2 &= - f(r) dt^2 + \frac{1}{f(r)} dr^2 + r^2 \big( H_e H_m \big)^h d S_2^2 \,, \\
f(r) &= \big( H_e H_m \big)^{-h} \Big( 1 - \frac{c}{4 \pi r} \Big) \,, \\
e^{-2 \lambda \phi} &= \Big( \frac{H_e}{H_m} \Big)^{2 - h} \,, \\
F &= \frac{q}{r^2} H_e^{-2} H_m^{2 - 2h} dt \wedge dr + p \, \text{sin} (\theta) d\theta \wedge d \varphi \,,
\end{split}
\label{eq:EMdBH}
\end{equation}
with $q,p$ electric and magnetic charges, respectively.
These solutions are well-defined for the coordinate values
\begin{align}
r \in \big( \tfrac{c}{4 \pi} \,, \infty\big) \,,
\label{eq:rrange}
\end{align}
where the extremality parameter ${c}$ defines the location of the outer horizon
\begin{align}
r_+ = \tfrac{c}{4 \pi} \,.
\end{align}
(The inner horizon is at $r_-=0$.)
The two undetermined functions $H_i (r)$, $i=e,m$, are solutions to the differential equation
\begin{align}
\label{diffH}
r^2 \frac{d}{d r} \bigg( r^2 \Big( 1 - \frac{c}{4 \pi r} \Big) \frac{H'_e (r)}{H_e (r)} \bigg) = - \frac{q^2}{h} H_e^{-2} H_m^{2 - 2h} \,,
\end{align}
and similarly for $e \leftrightarrow m$, $q \leftrightarrow p$. They are subject to the boundary conditions
\begin{align}
H_i (r) \underset{r \rightarrow \infty}{\longrightarrow} 1 \,, \quad \text{and} \quad  H_i (r) > 0 \,, \quad \text{for} \enspace r \rightarrow r_+ \,,
\end{align}
guaranteeing that the solutions are well-behaved in the region \eqref{eq:rrange}. These differential equations do not have a closed form for solutions at arbitrary value of $\lambda$. Nevertheless, solutions can still be parametrized as a power-series
\begin{align}
H_i (r) = 1 + \frac{\xi^{(1)}_i}{r} + \frac{\xi^{(2)}_i}{r^2} + \dots \,, \qquad i=e,m,
\label{eq:Hexpa}
\end{align}
in terms of constants $\xi^{(j)}_i$, and they converge within \eqref{eq:rrange}. Moreover, the system admits an integral of motion, which in the $r\to \infty$ limit reads
\begin{equation}
\label{1stint}
(\xi_e^{(1)})^2 + (\xi_m^{(1)})^2 + 2(h-1) \xi_e^{(1)} \xi_m^{(1)} + \tfrac{c}{4\pi} (\xi_e^{(1)}+\xi_m^{(1)}) - h^{-1}(q^2 + p^2) = 0,
\end{equation}
and it can be solved explicitly for some interesting values of the dilaton coupling $h$.

The above solutions approach extremality in the limit $c \rightarrow 0$. The entropy and temperature of the corresponding dilatonic black holes are given by
\begin{equation}\boxed{
\begin{split}
{\cal T} &= \tfrac{1}{{c}} \big( H_e \big( \tfrac{c}{4 \pi}\big) H_m \big( \tfrac{c}{4 \pi}\big) \big)^{-h} \,, \\
{\cal S} &= \tfrac{1}{2} {c}^2 \big( H_e \big( \tfrac{c}{4 \pi}\big) H_m \big( \tfrac{c}{4 \pi}\big) \big)^h \,.
\end{split}}
\label{eq:TSEMd}
\end{equation}
From which one again finds the extremality parameter
\begin{align}
2 {\cal S} {\cal T} = c \,.
\end{align}
The mass of the black hole is given by
\begin{align}
M =  c + 4 \pi h \big(\xi^{(1)}_e + \xi^{(1)}_m \big) \,,
\end{align}
which can be obtained from the metric in the limit of large $r$.

\subsection{Universal behavior in the extremal limit}
\label{subsec:univ}

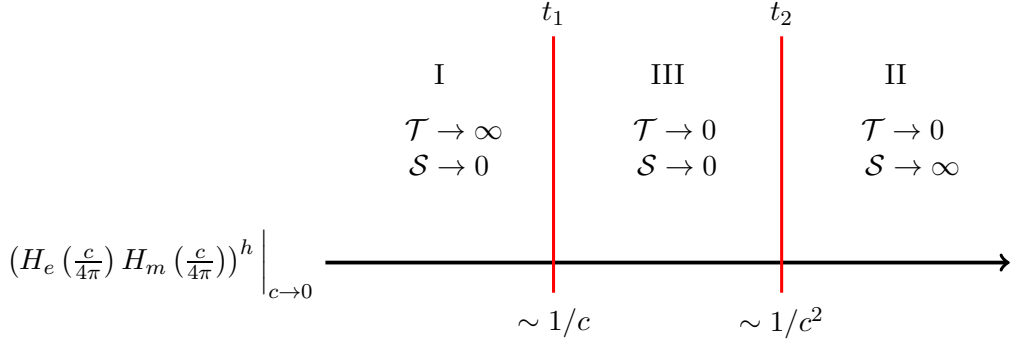
\begin{figure}
\begin{center}
\begin{tikzpicture}[scale=1.]
  \draw[->, line width=.5mm] (0, 0) node[left] {$\left(H_e \left(\frac{c}{4\pi}\right) H_m\left(\frac{c}{4\pi}\right)\right)^h\bigg|_{c\to 0}$} -- (9, 0) node[below] {};
  \draw[line width = .4mm, color=red] (3,-.4)--(3,3);
    \draw[line width = .4mm, color=red] (6,-.4)--(6,3);
  \node () at (3,-.8) {$\sim 1/c$};
    \node () at (3,3.3) {$t_1$};
       \node () at (6,3.3) {$t_2$};
  \node () at (6,-.8) {$\sim 1/c^2$};
  \node () at (1.5,1.5) {$\begin{array}{c}\, \, \,\  \,{\cal T}\to \infty \\ \  \  \mathcal{S} \to 0\end{array}$};
  \node () at (4.5,1.5) {$\begin{array}{c} \ \, {\cal T} \to 0 \\ \ \ \mathcal{S} \to 0\end{array}$};
    \node () at (7.5,1.5) {$\begin{array}{c} \ \, {\cal T}\to 0 \\ \ \ \,\, \, \mathcal{S} \to \infty\end{array}$};
    \node () at (1.5,2.5) {I};
    \node () at (4.5,2.5) {III};
    \node () at (7.5,2.5) {II};
\end{tikzpicture}
\end{center}
\caption{Depending on the behavior of $H_e(r) H_m (r)$ for $r \rightarrow 0$, there are three generic regions of the black hole thermodynamics, which are separated by threshold values.}
\label{fig:para_behav}
\end{figure}

Before analyzing some specific models, we want to explore the universal behavior of the above solutions in the extremal limit, $c \rightarrow 0$. 
First, we notice that in this limit the product $\mathcal{S} \mathcal{T}$ needs to vanish. The asymptotic behavior for ${\cal T}$ and $\mathcal{S}$  individually depends on the functions $H_i (r)$ in the vicinity of the outer horizon as well as the dilaton coupling $h$. With \eqref{eq:TSEMd}, one finds for $c \rightarrow 0$ the following possibilities.
\begin{equation}
\begin{split}
{\cal T} \rightarrow 0 \, : \, \text{if}& \quad H_e \big( \tfrac{c}{4 \pi} \big) H_m \big( \tfrac{c}{4 \pi} \big) \rightarrow \infty \quad \text{faster than} \quad {c}^{- \frac{1}{h}} \,, \\
{\cal T} \rightarrow \infty \, : \, \text{if}& \quad H_e \big( \tfrac{c}{4 \pi} \big) H_m \big( \tfrac{c}{4 \pi} \big) \rightarrow \infty \quad \text{slower than} \quad {c}^{- \frac{1}{h}} \,, \\
\mathcal{S} \rightarrow \infty \, :\, \text{if}& \quad  H_e \big( \tfrac{c}{4 \pi} \big) H_m \big( \tfrac{c}{4 \pi} \big) \rightarrow \infty \quad \text{faster than} \quad {c}^{- \frac{2}{h}} \,, \\
\mathcal{S} \rightarrow 0 \, :\, \text{if}& \quad  H_e \big( \tfrac{c}{4 \pi} \big) H_m \big( \tfrac{c}{4 \pi} \big) \rightarrow \infty \quad \text{slower than} \quad {c}^{- \frac{2}{h}} \,.
\end{split}
\end{equation}
This identifies three generic regions, I, II, and III, separated by threshold values $t_1$ and $t_2$ which allow for finite temperature and entropy, respectively. They are depicted in Figure \ref{fig:para_behav}.

In the parameter region I, the extremal limit leads to a state at infinite temperature with vanishing entropy. This is a small black hole limit. The specific realization of these states likely needs a UV complete description of the model and might be related to particle states as discussed in \cite{Holzhey:1991bx}. At the threshold $t_1$, the entropy still vanishes in the extremal limit, but one can achieve finite temperatures by balancing the parameters in the setup. The region II leads to infinitely large extremal black holes with vanishing temperature. These solutions are a priori not problematic from the viewpoint of the low energy effective description. In the parameter region III, both the temperature and the entropy of the extremal state vanishes. This is another small black hole limit. These charged extremal black holes of vanishing size are problematic in view of entropy arguments as those discussed in \cite{Hamada:2021yxy}. Once more, a UV description is necessary, whose higher curvature corrections induce a small but finite size of the extremal black holes. The appearance of small black holes in the effective theory can therefore be used to indicate that higher order effects become important. At the threshold $t_2$, the extremal black holes have vanishing temperature but a non-vanishing horizon area and entropy and thus they differ from the problematic small black holes of region III. A priori these black holes seem to be well-behaved in the effective field theory description. However, it has been argued that the semiclassical description of their thermodynamical properties can receive large corrections in the non-supersymmetric setup, see e.g.~\cite{Preskill:1991tb, Maldacena:1998uz, Page:2000dk, Heydeman:2020hhw} and references therein. 

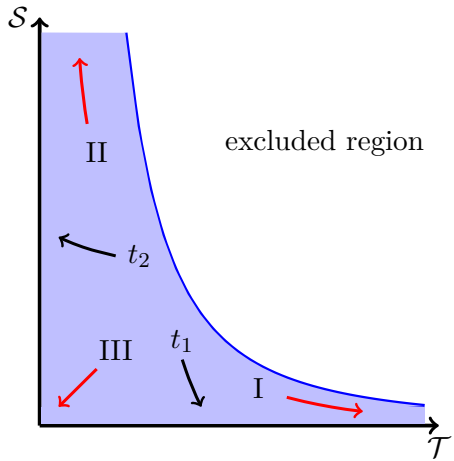
\begin{figure}
\begin{center}
\begin{tikzpicture}[scale=1.25]
  \draw[->, line width=.5mm] (0, 0) -- (4.2, 0) node[below] {${\cal T}$};
  \draw[->,line width=.5mm] (0, 0) -- (0,4.33) node[left] {${\cal S}$};
\draw[scale=1.5,domain=0.6:2.7,smooth, variable=\x,blue, thick,line width=.6mm] plot ({\x},{1/((\x)*(\x))});
\fill [blue!25, domain=0.1:4.2, variable=\x]
  (0.02, 4.17)
  -- plot[scale=1.5,domain=0.6:2.7] ({\x},{1/((\x)*(\x))})
  -- (0.02, 0.02);
  \fill [blue!25] (0.02,0.02) rectangle (4.05,0.2); 
  \node () at (3,3) {excluded region};
\draw[->,line width=.4mm,rounded corners=10pt, color=red] (2.6,.3)--(3,.2)--(3.4,.15);
\draw[->,line width=.4mm,rounded corners=10pt,color=red] (.6,.6)--(0.2,0.2);
\draw[->,line width=.4mm,rounded corners=10pt,color=red] (.5,3.2)--(.45,3.5)--(0.42,3.9);
\node () at (2.3,.4) {I};
\node () at (.8,.8) {III};
\node () at (.6,2.9) {II};
\draw[->,line width=.4mm,rounded corners=7pt,color=black] (.8,1.8)--(.4,1.9)--(0.2,2);
\draw[->,line width=.4mm,rounded corners=7pt,color=black] (1.5,.7)--(1.6,.4)--(1.7,.2);
\node () at (1.5,.9) {$t_1$};
\node () at (1.05,1.8) {$t_2$};
\end{tikzpicture}
\end{center}
\caption{Representation of the limits of Figure \ref{fig:para_behav} in the ${\cal T}- \mathcal{S}$ plane. The red lines correspond to the three different generic regions in Figure  \ref{fig:para_behav}, while the black lines correspond to the threshold values for which the function $(H_e H_m)^h$ vanishes as $\sim 1/c$ and $\sim 1/c^2$ respectively. In particular, the limit $t_1$ leads to ${\cal T} \text{ finite}$, $\mathcal{S} \to 0$, while $t_2$ leads to ${\cal T} \to 0$, $\mathcal{S} \text{ finite}$.}
\label{fig:BHpopulationandlimits}
\end{figure}

To illustrate the different universal regions, we refer to the diagram of possible charged black holes in Figure \ref{fig:BHpopulationandlimits}.
The three  generic limits I, II, and III are described by asymptotic points on the axes, whereas the threshold values potentially populate the axes as well. Next, we turn to the analysis of some special cases where the functions $H_i$ can be determined explicitly.

\subsection{Single-charged solutions}

EMd black holes with either electric charge $q$ or magnetic charge $p$ are arguably the simplest example and thus we discuss them first, before turning to the analysis of dyonic ones. Indeed, in this case one can find explicit solutions for $H_i$ for arbitrary dilaton coupling. 

We start from the case with magnetic charges only, for which one finds the solution
\begin{align}
H_e (r) = 1 \,, \quad H_m (r) = 1 + \frac{\xi_m}{r} \,,
\end{align}
with \eqref{1stint} reducing to
\begin{align}
h \xi_m \big(\xi_m +  \tfrac{c}{4 \pi} \big) = p^2 \,.
\end{align}
With this, we can derive the asymptotic behaviour for the extremal limit $c \rightarrow 0$: 
\begin{align}
H_e \big(\tfrac{c}{4 \pi} \big) H_m \big( \tfrac{c}{4 \pi} \big) \sim \tfrac{1}{{c}} \,.
\end{align} 
We see that the temperature goes to zero for $h > 1$ and the entropy vanishes in the extremal limit for $h < 2$. This means that for a non-trivial dilaton coupling, $\lambda > 0$, the extremal limit for a single-charged black hole always leads to a small black hole. The case at the $t_2$ threshold, $h = 2$ ($\lambda=0$), corresponds to a decoupling of the dilaton and reproduces the results for magnetically charged RN solutions, for which the entropy remains finite for $c \rightarrow 0$ and is given by
\begin{align}
\mathcal{S} = 4 \pi^2 p^2 \,, \quad \text{for} \enspace h = 2  \quad \text{($\lambda=0$)}\,.
\end{align}
This is the first example we analyse of a black hole solution coupled to a scalar field. Thus, we are interested in studying the profile of the dilaton. For $\lambda > 0$, one has (see \eqref{eq:EMdBH})
\begin{align}
\left(g(\phi)\right)^{-2} \equiv e^{- 2 \lambda \phi} = \left( 1 + \frac{\xi_m}{r} \right)^{h - 2} \,,
\end{align}
where $g$ denotes the electric gauge coupling. The above quantity, which is in fact the (squared) magnetic gauge coupling, vanishes at the outer horizon $r_+ = \tfrac{c}{4\pi}$ in the extremal limit (for $h < 2$). This indicates that the dilaton diverges
\begin{align}\label{eq:dilat-sing-charge}
\phi \rightarrow \infty \,.
\end{align}
Since the asymptotic value of the dilaton in this solution is fixed to $\phi |_{r \rightarrow \infty} = 0$ , we see that for purely magnetically charged extremal black holes the dilaton traverses an infinite field distance and, at the same time, a global symmetry is restored. This infinite field distance (or global symmetry restoration) gives an alternative perspective in terms of the SDC on our intuition that the small black hole limit is outside the validity of the effective description, which should be related to a light tower of states close to the black hole horizon.

For the coupling $h=1$, the single charge black hole realizes the threshold case $t_1$ of Figure \ref{fig:para_behav}, see also \cite{Holzhey:1991bx}. This corresponds to the charged dilatonic black hole in string theory, with only one charge. In the extremal case the solution has zero horizon but finite temperature. In addition to the singular behavior of the dilaton \eqref{eq:dilat-sing-charge}, in the $t_1$ limit we also have a singular geometry due to the vanishing size of the horizon. Again we expect quantum gravity corrections to take over and smooth out the classical singularity.

The case of purely electrically charged black holes can be deduced from the electric-magnetic duality of the action, under which
\begin{align}
\left( \begin{array}{c} F_{\mu\nu} \\ G_{\mu\nu} \end{array} \right) \rightarrow \left( \begin{array}{c} G_{\mu\nu} \\ -F_{\mu\nu} \end{array} \right) \,, \quad \phi \rightarrow - \phi,
\end{align}
where $G_{\mu \nu} = \frac{\delta\mathcal{L}}{\delta F_{\mu\nu}}=\tfrac{1}{2} e^{-2 \lambda \phi} \epsilon_{\mu \nu \rho \sigma} F^{\rho \sigma}$. Once more, the entropy vanishes in the extremal case unless $\lambda = 0$. Since the dilaton changes sign one now has
\begin{align}
\phi \rightarrow - \infty \,,
\end{align}
at the outer horizon and again the traversed field distance in the black hole background is infinite. A recent discussion on this class of black holes in the Swampland context can be found in  \cite{Hamada:2021yxy}. In particular, it is argued that the vanishing size of extremal EMd black holes with charge leads to a violation of entropy bounds. Additionally, one finds that in these situations the dilaton traverses infinite distance. The SDC is thus assumed to avoid such a problematic situation by invalidating the effective theory with the appearance of a tower of light states. RN black holes evade this conclusion since they still have a finite size in the extremal limit. Thus, the role of the scalars (dilaton) in this class of model is crucial for understanding properties of quantum gravity.

To summarize, we see that for non-trivial dilaton coupling the limit of extremal single-charge solutions is problematic since it is described by the parameter region II and is beyond the regime of validity for the Einstein-Maxwell-dilaton effective theory. The usual expectation is that the full UV-complete theory contains higher derivative corrections that lead to a finite horizon size. Note that in all of the above cases the problematic regime ${\cal S} \rightarrow 0$ is accompanied by an infinite field distance traversed by the dilaton field from spatial infinity to the outer horizon relating it to the predictions of the SDC.

\subsection{Dyonic solutions}
\label{subsec:dyon}

We now turn to solutions with both electric and magnetic charges, dyonic black holes. Unfortunately, these cannot be discussed in full generality for arbitrary $h$ with one exception, for which the electric and magnetic charges are equal
\begin{align}
\label{p=q}
p = q \,.
\end{align}
The corresponding solution is given by
\begin{align}
H_e (r) = H_m (r) = \Big( 1 + \frac{\xi}{r} \Big)^{1/h} \,, \quad \xi \big(\xi +  \tfrac{c}{4 \pi} \big) = q^2 \,,
\end{align}
leading to the asymptotic behavior in the extremal limit
\begin{align}
H_e \big( \tfrac{c}{4\pi} \big) H_m \big( \tfrac{c}{4\pi} \big) \sim {c}^{-2/h} \,,
\end{align}
located precisely at the threshold  $t_2$.
In fact, in the extremal limit one finds that the entropy
\begin{align}
\mathcal{S} = 8 \pi^2 q^2 \,,
\end{align}
is finite and independent of $h$. At the same time the dilaton profile is constant. It therefore seems that for these black holes one can reach the extremal limit within the effective description and correspondingly one does not traverse an infinite distance in the moduli space of the dilaton field. Moreover, also in the infinite entropy limit, $q^2\rightarrow \infty$, $e^{-2\lambda\phi}$ stays constant and one again does not traverse an infinite distance in the moduli space of $\phi$. Note, however, that especially in non-supersymmetric cases, these black holes might receive large quantum corrections affecting their thermodynamical properties \cite{Preskill:1991tb, Maldacena:1998uz, Page:2000dk, Heydeman:2020hhw}.

For the analysis at arbitrary charges, we focus on two different interesting cases for which explicit solutions of $H_i (r)$ are known (the case $h = 2$ is fully covered by the considerations above):
\begin{itemize}
	\item{$h = 1$: Low-energy effective action of heterotic string theory}
	\item{$h = \tfrac{1}{2}$: Kaluza-Klein theory, with the dilaton as radion field}
\end{itemize}
We will see that in these examples the extremal limit for dyonic black holes corresponds to the threshold $t_2$ and thus the entropy remains finite and the dilaton does not traverse to infinite distance.

\subsubsection*{String theory coupling ($h = 1$, $\lambda=1/\sqrt2$)}

For the typical string theory coupling of the dilaton, i.e., $h = 1$, one finds the following solution
\begin{align}
H_i = 1 + \frac{\xi_i}{r} \,, \quad \xi_i \big(\xi_i + \tfrac{c}{4\pi} \big) = q_i^2 \,,
\end{align}
where $q_e = q$ and $q_m = p$. 

When both charges are nonzero, in the extremal limit one recovers
\begin{align}
H_e \big( \tfrac{c}{4\pi} \big) H_m\big( \tfrac{c}{4\pi} \big) \sim {c}^{-2} \,. 
\end{align}
Again, this is at the threshold $t_2$, where
\begin{align}
{\cal S} \enspace \underset{{c} \rightarrow 0}{\longrightarrow} \enspace 8 \pi^2  p q \,.
\end{align}
The dilaton field further has the asymptotic value
\begin{align}
e^{- 2 \lambda \phi} \Big|_{r_+} \enspace \underset{{c} \rightarrow 0}{\longrightarrow} \enspace \frac{q}{p} \,,
\end{align}
at the outer horizon. The field distance in moduli space is therefore finite, for finite $p$ and $q$ and the extremal limit seems to be within the validity of the effective description.

When only one of the charges is nonzero, we recover the $t_1$ threshold states discussed above.

\subsubsection*{Kaluza-Klein coupling ($h = \tfrac{1}{2}$, $\lambda=\sqrt{3\over2}$)}

For the Kaluza-Klein value of the dilaton coupling one has the solutions
\begin{align}
H_i (r) = 1 + \frac{\xi_i^{(1)}}{r} + \frac{\xi_i^{(2)}}{r^2} \,,
\end{align}
with
\begin{align}
2 q_i^2 = \frac{\xi_i^{(1)} \big( \xi_i^{(1)} + \tfrac{c}{4 \pi} \big) \big( \xi_i^{(1)} + \tfrac{c}{2 \pi} \big)}{\xi_e^{(1)} + \xi_m^{(1)} + \tfrac{c}{2\pi}} \,, \quad \xi_i^{(2)} = \frac{\xi_e^{(1)} \xi_m^{(1)} \big( \xi_i^{(1)} + \tfrac{c}{4 \pi} \big)}{2 \big( \xi_e^{(1)} + \xi_m^{(1)} + \tfrac{c}{2 \pi} \big)} \,.
\end{align}
and the extremal limit is characterized by
\begin{align}
H_e \big( \tfrac{c}{4\pi} \big) H_m \big( \tfrac{c}{4\pi} \big)\sim {c}^{-4} \,,
\end{align}
which is again at the threshold $t_2$.
Once more the entropy in the extremal limit is determined by the electric and magnetic charges
\begin{align}
{\cal S} \enspace \underset{{c} \rightarrow 0}{\longrightarrow} \enspace 8 \pi^2  p q \,,
\end{align}
as is the asymptotic value of the dilaton field
\begin{align}
e^{- 2 \lambda \phi} \Big|_{r_+} \enspace \underset{{c} \rightarrow 0}{\longrightarrow} \enspace \frac{q}{p} \,.
\end{align}
Therefore, for finite values of the charges the extremal black holes have a finite entropy and do not lead to an infinite field distance in the moduli space of the dilaton.

\subsection{Infinite entropy at threshold}

We have seen that the dyonic black holes are realized at the threshold value $t_2$ of Figure \ref{fig:BHpopulationandlimits}. This implies that for finite charges their entropy remains non-vanishing and finite in the extremal limit. There are, however, limits in which one sends the charges to infinity that also lead to infinite entropy. These limits further affect the dilaton field profile, which we want to analyze next.

\subsubsection*{Infinite entropy and infinite scalar field limit}

For the first limit we consider dyonic black holes with electric and magnetic charges given by $q$ and $p$, respectively. The entropy in the cases analyzed above is proportional to their product. Therefore, one can generate infinite entropy by keeping one of the charges fixed and sending the other one to infinity. This generates a hierarchy in the two charges and induces a large field limit for the dilaton, since its value diverges at the outer horizon. Concretely, for these two cases the entropy can be expressed in terms of the scalar field as
\begin{eqnarray}
p \text{ fixed}:\quad&{~}&{\cal S} \enspace \underset{{c} \rightarrow 0}{\longrightarrow} \enspace 8 \pi^2  p^2 e^{-2 \lambda \phi} \Big|_{r_+}
\quad {\rm or}\nonumber\\
q \text{ fixed}:\quad&{~}&{\cal S} \enspace \underset{{c} \rightarrow 0}{\longrightarrow} \enspace  8 \pi^2  q^2 e^{ 2 \lambda \phi} \Big|_{r_+} 
\end{eqnarray}
So for large entropy and fixed magnetic  or fixed electric charge we get 
\begin{equation}
\begin{split}
p \text{ fixed}:\quad{\cal S} \rightarrow \infty:& \quad \phi \rightarrow - \infty \,, \\
q \text{ fixed}:\quad{\cal S} \rightarrow \infty:& \quad \phi \rightarrow \infty \,.
\end{split}
\end{equation}
We see that the infinite entropy limit corresponds to an infinite distance limit in the scalar field moduli space, again, demonstrating the parallels with the SDC and its predictions\footnote{Note that if the spacetime region, at which the infinite distance limit is achieved, has vanishing volume, one in general does not expect an infinite tower of states. An example for this situation is the Taub-NUT geometry. This often seems to be associated with the fact that these backgrounds are non-trivial fibrations, in which the charge labelling the infinite tower, such as the KK charge, is not conserved. In our case, we consider geometries which are warped products and, in particular, are trivial fibrations. Moreover, once corrections are taken into account, a finite horizon is generated. We therefore expect the SDC to be applicable in our set-up. We thank the referee for pointing out this issue.}. Moreover, these two limits are dual to each other, in the sense that the infinite entropy limit either corresponds to very weak or very strong $U(1)$ gauge coupling.
In a string theory embedding, small dilatonic black holes should correspond to regions of strong/weak string coupling limits ($g_s\rightarrow \infty/0$) and then to an infinite tower of light strings \cite{Lee:2019wij}.

\subsubsection*{Infinite entropy and finite scalar field limit}

Next, we consider the case, where the electric and magnetic charges scale in the same way, i.e.,
\begin{align}
p \sim {q} \,.
\end{align}
This case was already discussed in Section (\ref{subsec:dyon}). The value of the dilaton field stays finite in the infinite entropy limit:
\begin{equation}
\begin{split}
{\cal S} \rightarrow \infty: \quad e^{ 2 \lambda \phi} \rightarrow  {\rm const}\,.
\end{split}
\end{equation}
So the infinite entropy limit corresponds to a finite field distance in the moduli space of the scalar field. We will come back to these ``scaling'' solutions in the context of $\mathcal{N} = 2$ supergravity.

\subsubsection*{Finite entropy and infinite scalar field limit}

The dyonic black hole solutions further allow for a parametric solution with 
\begin{align}
p \sim \frac{1}{q} \,,
\end{align}
which for $q \rightarrow \infty$, seems to lead to a finite entropy black hole in the extremal limit. At the same time the value of the dilaton at the outer horizon takes one of its singular values
\begin{equation}
\begin{split}
q \rightarrow \infty:& \quad \phi \rightarrow - \infty \,, \\
p \rightarrow \infty:& \quad \phi \rightarrow \infty \,.
\end{split}
\end{equation}
Therefore they seem to correspond to finite entropy black holes, which nevertheless induce an infinite field distance in the moduli space of the dilaton and are problematic from an effective field theory point of view. 

\subsection{Infinite temperature limit}

The infinite temperature limit in the extremal case is captured already by the discussion in Section \ref{subsec:univ} and corresponds to the asymptotic behavior in the parameter region I, i.e., 
\begin{align}
H_e \big( \tfrac{c}{4\pi} \big) H_m \big( \tfrac{c}{4\pi} \big) \rightarrow \infty \quad \text{slower than} \quad {c}^{- \frac{1}{h}} \,. 
\end{align}
For non-vanishing ${c}$ we find that ${\cal T} \rightarrow \infty$ demands
\begin{align}
H_e \big( \tfrac{c}{4\pi} \big) H_m \big( \tfrac{c}{4\pi} \big) \rightarrow 0 \,.
\end{align}
Since the boundary conditions are such that $H_i (r) > 0$ for $r \rightarrow \tfrac{c}{4\pi}$, we see that this is not possible. To be more precise for finite ${c}$ this demands that some of the terms in expansion \eqref{eq:Hexpa} are negative. However, these will dominate in the extremal limit and violate the boundary condition. Therefore, all infinite temperature limits of the black holes above necessarily are extremal and therefore captured by the asymptotic behavior above. This can also be seen in the Figure \ref{fig:BHpopulation} and \ref{fig:BHpopulationandlimits}, where in the limit ${\cal T} \rightarrow \infty$ one approaches the axis, i.e., the extremal black holes, and all finite values of $c$ exit the allowed region at a certain point. 

\vskip0.4cm \noindent
We see that already in the simplest model of gravity coupled to a scalar field, Einstein-Maxwell-dilaton theory, we can relate many of the thermodynamic limits to infinite field ranges of the dilaton field in the black hole background. The SDC then predicts a tower of light states to appear in the vicinity of the black holes leading to large corrections and necessitate an understanding within a more fundamental description. In order to explicitly study these light states we therefore turn to an effective theory closely related to string theory, namely $\mathcal{N} = 2$ supergravity.

\section{Multi-charge black holes in ${\cal N}=2$ supergravity}
\label{SG-blackholes}

The behavior of black holes solutions across the parameter space explored in the previous section can also be found for more general, multi-charge black holes in four-dimensional $\mathcal{N}=2$ supergravity coupled to vector multiplets.
The main example we will concentrate on is the ${\it STU}$ model arising from type IIA compactifications, but our analysis can be easily adapted to different Calabi-Yau spaces or dualized to type IIB.\footnote{An exhaustive review on ${\cal N}=2$ supergravity in the context of black holes is \cite{Andrianopoli:2006ub}. For BPS black holes and their relation to string theory compactified on Calabi-Yau we refer to \cite{Behrndt:1996jn,Greene:1996cy,Bohm:1999uk,Denef:1999idt,Shmakova:1996nz}. See also \cite{Cvetic:1995uj,Cvetic:1995bj}  for $\mathcal{N}=4$ black holes.}

The four-dimensional effective $\mathcal{N}=2$ supergravity theory arising from type IIA string theory compactified on a Calabi-Yau threefold $Y$ is specified by a prepotential 
\begin{align}
F(X^\Lambda)=\frac16 d_{ijk}\frac{X^{i}X^{j}X^{k}}{X^0} \ ,
\end{align}
where $X^\Lambda=(X^0,X^i)$, and the indices $i,j,k$ run over $1,..,h^{1,1} \equiv n_V$. 
The bosonic sector contains the graviton, $n_V+1$ vectors and $n_V \equiv h^{1,1}$ complex scalars $z^i$ spanning a special K\"ahler manifold. A convenient formulation of special geometry is in terms of projective coordinates $X^\Lambda(z^i)$ which, together with $F_\Lambda = \frac{\partial F}{ \partial X^\Lambda} $ (valid only if a prepotential exists), form holomorphic symplectic sections $(X^\Lambda,F_\Lambda)$. They arise from a choice of symplectic basis of 2- and 4-cycles $\{A^\Lambda\,, B_\Lambda\}$  on $Y$. 
Finally, $d_{ijk}$ gives the triple intersection numbers of the Calabi-Yau 4-cycles.

Four-dimensional black holes are solutions of the supergravity action and, upon a constraint on the parameters, at zero temperature they can preserve supersymmetry. We choose a microscopic configuration obtained from 10d as a superposition of $h^{1,1}$ D4-branes, each of them wrapped $p_i$-times around the corresponding 4-cycles in $Y$, times a stack of $q_0$ D0-branes.
In four dimensions, they yield a black hole with electric charge $q_0$ and magnetic charges $p_i$ with respect to the abelian ${h^{1,1}+1}$ gauge fields. Analogous results to those we present can be dualized to the D2-D6 setup, where the black hole has instead one magnetic $p^0$ and electric $q_i$ charges.  We refer the reader to Appendix \ref{appendix-supergravity} for more details on four-dimensional $\mathcal{N}=2$ supergravity and for definitions of the quantities further introduced in this section. 

The explicit setup we will focus on is the so-called ${\it STU}$ model, corresponding to the case where $d_{123}=1$ and zero otherwise. We have three scalars $z^1=S\,,z^2=T\,,z^3=U$, defined as $z^i=\frac{X^i}{X^0}$, which enter the action as a nonlinear sigma model with target space $\left[SU(1,1)/U(1)\right]^3$. The ${\it STU}$ model can be interpreted as a truncation of $\mathcal{N}=8$ supergravity arising from type IIA compactified on $T^6$, in which case the imaginary parts of the three scalar fields parametrize the volume moduli of three two-tori. It can also arise as a compactification of heterotic string theory on $T^2\times K_3$, which is dual to type IIA compactified on a CY with $h^{1,1}=3$ \cite{Behrndt:1996jn}.
In the limits $S,T,U\rightarrow 0,\infty$ (neglecting the axions), the volumes of these two-tori will either shrink to zero or grow to infinite sizes. 

It is known that these wo limits are at infinite distance in the internal moduli space.  In fact the overall volume of the Calabi-Yau is given by  
\begin{equation} \label{volumeCY3}
\mathcal{V} = -i z^1 z^2 z^3 =\,-iSTU\,,
\end{equation}
where we implicitly set $\Re z^i=0$. 
The overall KK modes have masses (in string units)
\begin{equation}
m_{KK}= \left(\frac{1}{\mathcal{V}}\right)^{1/6}\, .
\end{equation}
In case also string winding modes are present, as for example on the torus $T^6$, their masses are given by
\begin{equation}
m_{wind}={\mathcal{V}}^{1/6}\, .
\end{equation}
Therefore, in the limits ${\mathcal{V}}\rightarrow \infty$ or  ${\mathcal{V}}\rightarrow 0$, either the tower of KK or winding modes (if present) become massless, in agreement with the infinite distance in the internal moduli spaces.
Specifically, the moduli space distance in the limit of large or small volume behaves as
\begin{equation}
\Delta_{\cal V}\simeq |\log({\cal V})|\, .
\end{equation}
In the next sections, we will see how the internal volume is mapped to the entropy and the temperature of the non-extremal black hole solutions.

\subsection{Non-extremal black holes \label{sec:non-extremal}}

Black hole solutions with electric and magnetic charges at finite temperature can be described in supergravity by the metric\footnote{In the present Section \ref{SG-blackholes} we set $G_N=1$, as we are referring to the supergravity black hole literature which uses these conventions. This will affect the Schwarzschild curve in the ${\cal T}-{\cal S}$ plane defined as ${\cal S}{\cal T}^2=\frac{1}{16\pi G_N}$, which here will appear as ${\cal S}{\cal T}^2=\frac{1}{16\pi}$. One can restore the explicit $G_N$ dependence by setting $c \to G_N c$ everywhere in the solution presented in this Section;  thermodynamic quantities can be compared with those in the previous sections by transforming \eqref{S-T-BH} as
  \begin{align}
    {\cal S}(c)&\to \frac1{G_N}{\cal S}(G_N c) \ , 
    \qquad \qquad 
    {\cal T}(c) \to {\cal T}(G_N c) \ ,
  \end{align}
and then chosing $G_N=\frac{1}{8\pi}$.
} 
\begin{align}
\label{metric}
ds^2&=-e^{2U}dt^2+e^{-2U}\left[ \frac{c^4 d\rho^2}{\sinh^4(c\rho)} + \frac{c^2}{\sinh^2(c\rho)} \left(d\theta^2+\sin^2\theta d\phi^2\right) \right] \ ,
\end{align}
where the warp factor $e^{2U}$ represents the coupling to scalar fields \cite{Breitenlohner:1987dg}. This metric interpolates between the extremal ($c=0$) and non-extremal $(c> 0)$ geometries.
The black hole is characterized by a mass $M$ and by electric and magnetic charges, $q_\Lambda$ and $p^\Lambda$, with $\Lambda \in \{0,1,..,n_V\}$. Due to the coupling to the scalars, $M$ is fixed by $c$, the black hole charges and also by the values of the scalar fields at asymptotic infinity. The latter determine the nature of the specific compactification and we decide to work with a fixed choice of moduli at infinity, i.e., we fix the low energy theory. Our conclusions will not depend on this specific choice, as long as we remain inside the moduli space.  Besides the metric, the black hole solution is specified also by the field strengths
\begin{align}
\label{Fmunu-spher-symm-4d}
F^\Lambda&=p^\Lambda \sin\theta \, d\theta \wedge d\phi - e^{2U}  (\mathcal{I}^{-1})^{\Lambda\Sigma}\left(q_\Sigma-\mathcal{R}_{\Sigma\Delta}p^\Delta\right) d\rho \wedge dt \ ,
\end{align}
where $\mathcal{I}_{\Lambda\Sigma}(z^i,\bar{z}^{\imath})\,,\mathcal{R}_{\Lambda\Sigma}(z^i,\bar{z}^{\imath})$ are scalar couplings to the gauge fields in the theory entering the action in front of the gauge kinetic and theta-term, respectively (see Appendix \ref{appendix-supergravity}).

The \emph{STU} model admits black holes with up to four electric and four magnetic charges \cite{Galli:2011fq,Bellucci:2008sv}. For simplicity, we work with a configuration with only nonvanishing $q_0,p^1,p^2,p^3$ charges, all assumed to be positive. The solution can be written in terms of four functions $I_0(\rho)\,, I^1(\rho)\,, I^2(\rho)\,, I^3(\rho)$, in such a way that the scalar fields are
\begin{equation}
\label{scalars-stu}
z^i=-i\frac{I_0I^i}{\sqrt{\mathcal{I}_4}} \ , \qquad  \mathcal{I}_4(I)=I_0 I^1I^2I^3 \ ,
\end{equation}
and the warp factor is
\begin{equation}
\label{warp-fact-nonextr}
e^{-2U}= e^{-2U_e+2c\rho}\ , \qquad \text{with} \qquad e^{-2U_e}=4\sqrt{\mathcal{I}_4}\ .
\end{equation}
The $I$-functions take the form
\begin{equation}\label{I-func}
I_0=a_0+b_0 e^{2c\rho} \ , \qquad \qquad  I^i=a^i+b^i e^{2c\rho} \ ,
\end{equation}
with
\begin{align}
\label{a0b0}
\begin{pmatrix} a_0 \\ b_0 \end{pmatrix} &=  \, \frac{1 }{8\sqrt2 L^0_{\infty}}
\left[1\pm \frac{1}{c}\sqrt{c^2+16q_0^2(L_{\infty}^0)^2}\right] \ ,  \\ \label{a1b1}
\begin{pmatrix}  a^i \\ b^i \end{pmatrix}  &= -   \, \frac{1}{8\sqrt2 M_{i\, \infty}}\left[1\pm\frac{1}{c}\sqrt{c^2+ 16 (p^i)^2 M^2_{i\,\infty} } \right]\,,
\end{align}
and
\begin{align}
\label{simpl-secs-infty}
L^0_\infty&= \frac{1}{2\sqrt2}\frac{1}{\sqrt{\lambda^1_\infty\lambda^2_\infty\lambda^3_\infty}}\ , \qquad \qquad  M_{i\,\infty }= -\frac{\sqrt{\lambda^1_\infty\lambda^2_\infty\lambda^3_\infty}}{2\sqrt2 \lambda^i_\infty}\ ,
\end{align}
where we denoted  $z^i_\infty=-i\lambda^i_\infty$ the values of the scalar fields at infinity, that will be kept fixed in all our analysis. Notice that the horizon is located at $\rho\to-\infty$, so the value of the scalars at the horizon are in principle dependent on the charges, the moduli at infinity, and on $c$. 

From the expression of the metric in \eqref{metric}, we derive that the black hole entropy and temperature are
\begin{equation}
\label{S-T-BH}
\boxed{
\begin{split}
{\cal S}=& 16 \pi \sqrt{a_0a^1a^2a^3} \, c^2 \ , 
\\
{\cal T}=& \frac{\kappa}{2\pi} = \frac{1}{2\pi} \left[ \frac12 \left(\frac{\sinh(c\rho)^2}{c^2}\frac{d e^{2U(\rho)}}{d\rho}\right)
\right]_{\rho\to-\infty} = \frac{1}{32\pi}\frac{1}{c\sqrt{a_0a^1a^2a^3}}\, ,
\end{split} }
\end{equation} 
and the extremality parameter is
\begin{equation}
\label{nonextremalityrelation}
c=2{\cal S}{\cal T} \ .
\end{equation}

This solution generalizes non-extremal black holes in the presence of scalar fields running from asymptotic infinity to their values at the horizon. When the scalars are constant throughout the whole four-dimensional spacetime, the solution effectively reduces to a non-extremal Reissner-Nordstrom black hole. 

The volume of the Calabi-Yau for this solution is given by  
\begin{equation} 
\label{volumeCY3-stu}
\mathcal{V} =- i S T U = \frac{I_0^2}{\sqrt{\mathcal{I}_4}}\,,
\end{equation}
with the $I$-functions defined in \eqref{I-func}.

\subsection{Extremal BPS black holes}

Supersymmetric black holes are an important subset of the solutions of the previous section. Supersymmetry requires the solution to be extremal, meaning $c=0$, and thus \eqref{nonextremalityrelation} implies $\mathcal{T}=0$ for a regular horizon. Therefore, according to our classification of limits, we are dealing here with the threshold case $t_2$.

This class of solutions is described by a metric of the form\footnote{Notice that the coordinates $r$ and $\rho$ are related by $dr=-\frac{c^2}{\sinh(c\rho)^2}d\rho $, which in the extremal case simply reduces to $dr=-\frac{d\rho}{\rho^2}$.}
\begin{equation}
\label{dsextr1}
ds^2= -e^{2U} dt^2 + e^{-2U} ( dr^2 + r^2 d\Omega_{2}^2) \ .
\end{equation}
Not all possible choices of charges in the theory allow for a BPS solution, i.e., the extremal limit of a multi-charge black hole in supergravity may lead to non-BPS solutions with finite entropy. However, the non-extremal configuration of the \emph{STU} model with charges $q_0\,, p^1\,,p^2\,,p^3$  considered so far does reduce to a supersymmetric black hole for $c\to 0$.\footnote{We refer to \cite{Bellucci:2008sv} for a review of BPS and non-BPS black holes in the \emph{STU} model.} This BPS black hole is given by the scalar fields $z^i$ of \eqref{scalars-stu} and the warp factor $e^{-2U}\to e^{-2U_e}$ of \eqref{warp-fact-nonextr}, with
\begin{equation}
I_0= \frac{1}{4\sqrt2 L^0_\infty}-\frac{1}{\sqrt2} \frac{q_0}{r} \ ,\qquad  I^i= -\frac{1}{4\sqrt2 M_{i\,\infty}}-\frac{1}{\sqrt2}\frac{p^i}{r} \ .
\end{equation}

Being extremal, BPS black holes in $\mathcal{N}=2$ supergravity have a near horizon geometry of the form $AdS_2\times S^2$, with the anti-de Sitter radius equal to the sphere radius and proportional to the black hole entropy. These black holes are governed by the central charge function $Z(z^i,\bar z^{\bar \imath},q_0,p^i)$ defined in Appendix \ref{appendix-supergravity}, and the horizon corresponds to an attractor point for the flow of the scalar fields determined by $Z$ via the \emph{attractor equations} 
\begin{align}
&\partial_i Z(z^i_{h},\bar z^{\bar \imath}_{h},q_0,p^i)=0\ , \\
 &{\cal S}_{BPS}=\pi |Z(z^i_{h},\bar z^{\bar \imath}_{h},q_0,p^i)|=2\pi \sqrt{q_0p^1p^2p^3}\ .
\end{align}
For the BPS black hole of the $STU$ model with charges $q_0\,, p^1\,,p^2\,,p^3$, this yields
\begin{align}\label{BPS-sol}
z^i_{h}&=-i\frac{q_0 p^i}{\sqrt{q_0p^1p^2p^3}} \ , \\
 {\cal S}_{BPS}&=2\pi \sqrt{q_0p^1p^2p^3}\ ,
\end{align}
corresponding to the $c\to0$ limit of \eqref{S-T-BH}. We see that in the BPS case the scalars at the horizon are completely fixed by the charges, differently from the non-extremal solution where the horizon values still depend on the moduli at infinity, in addition to the extremality parameter $c$. The mass of the BPS black hole is also determined in terms of the central charge as $M~=~|Z(z^i_{h},\bar z^{\bar \imath}_{h},q_0,p^i)|$.

We will now explore the parameter space of this general class of $STU$ black holes in the ${\cal T}-{\cal S}$ space.

\subsection{Thermodynamic limits and tower of states}\label{thlimits}

In this section, we investigate the limits of large and small temperature or entropy for extremal and non-extremal black holes in supergravity, 
by studying the behaviour of the CY volume \eqref{volumeCY3}  at the horizon, whose divergence or singular behaviour may signal the appearance of infinite towers of states. We will specialize our discussion to the $STU$ black holes presented above, for concreteness, but our results can be easily generalized to other configurations.  

\subsubsection{Extremal black holes}\label{Sec.Extremal}

Let us first consider the BPS black hole case. The thermodynamic quantities in the extremal limit are, from \eqref{BPS-sol} 
\begin{equation}
\boxed{
\begin{split}
{\cal S}&=2\pi \sqrt{q_0 p^1 p^2 p^3} \ , \\
{\cal T} &=  0\,.
\end{split}}
\end{equation}
The volume (\ref{volumeCY3-stu}) at the horizon is thus given in terms of the charges as
\begin{equation}\boxed{
\mathcal{V}_h = \frac{(q_0)^{3/2}}{\sqrt{p^1 p^2 p^3}}\,.}
\end{equation}
The KK modes have masses 
\begin{equation}
m_{KK}= \left(\frac{1}{\mathcal{V}_h}\right)^{1/6}=\frac{(p^1 p^2 p^3)^{1/12}}{(q_0)^{1/4}}\,.
\end{equation}
Let us now consider the exchange of a small extremal black hole with a large extremal one, which can be realized by the inversion of the charges (and of the mass):
\begin{equation}
q_0\,\longleftrightarrow\,{1\over q_0}\,, \quad 
p^i\,\longleftrightarrow\,{1\over p^i} \quad 
\Longrightarrow
\quad
{\cal S}\,\longleftrightarrow\,{1\over{\cal S}}\, .
\end{equation}
This transformation also inverts the volume of the internal space, i.e.,
\begin{equation}
{\cal V}_h\,\longleftrightarrow\,{1\over{\cal V}_h}\, ,
\end{equation}
and exchanges the KK modes with the winding modes.
So this exchange of large and small extremal black holes is induced by a T-duality transformation on the internal moduli fields.

Now, let us see if the large or small entropy limits also induce a tower of light states and hence leads to an infinite distance limit in the moduli space. For this purpose, we will focus on two types of black holes, which will exhibit a kind of T-dual behavior with respect to each other:

\begin{itemize}
\item {\bf Type A}: Black holes with \emph{fixed electric charge $q_0$}, varying magnetic charges $p^i$ and varying mass $M$. The entropy will therefore be a function of the $p^i$ and the mass $M$.
\item {\bf Type B}: Black holes with \emph{fixed magnetic charges $p^i$}, varying electric charge $q_0$ and varying mass $M$. Then the entropy will be a function of $q_0$ and the mass $M$.
\end{itemize}

\noindent
For A-type black holes, the volume scales as a {\it negative} power of the entropy
\begin{equation}\label{Vextremalq}
\boxed{
\mathcal{V}_h = 2\pi\frac{q_0^2}{\cal S}\,.}
\end{equation}
This implies that for fixed electric charge $q_0$ the KK masses grow together with the entropy as
\begin{equation}
m_{KK}=\frac{1}{(2\pi)^{1/6}} \frac{{\cal S}^{1/6}}{q_0^{1/3}}\,
\end{equation}
and thus become light in the limit of small entropy. For toroidal compactification on $T^6$, there are also T-dual winding modes with masses of the order
\begin{equation}
m_{wind}=\mathcal{V}_h^{1/6}\simeq \frac{{q_0}^{1/3}}{{\cal S}^{1/6}}\,,
\end{equation}
which become light in the limit of large entropy.

\vskip0.2cm
On the other hand for the B-type black holes the volume scales as a {\it positive} power of the entropy
\begin{equation}\label{Vextremalp}\boxed{
\mathcal{V}_h = \frac{1}{(2\pi)^3}\frac{{\cal S}^3} {(p^1 p^2 p^3)^2}\,,}
\end{equation}
and the KK masses decrease in the large entropy limit as
\begin{equation}
m_{KK}=(2\pi)^{1/2} \frac{(p^1 p^2 p^3)^{1/3}}{{\cal S}^{1/2}}\,.
\end{equation}
In case there are T-dual winding modes they have masses 
\begin{equation}
m_{wind}\simeq\frac{{\cal S}^{1/2}}{(p^1 p^2 p^3)^{1/3}}\,
\end{equation}
and become light in the limit of small entropy.

\vskip0.2cm
The  A- and B-type black holes, i.e.~fixed $q_0$ or $p^i$, behave T-dual to each other, since sending ${\cal S}\rightarrow0$ or $\mathcal{S}\to\infty$ implies opposite behavior for the volume, i.e.~$\mathcal{V}_h\rightarrow 0$ or $\mathcal{V}_h\rightarrow\infty$, and in turn for the KK masses, i.e., $m_{KK}\rightarrow\infty$ or $m_{KK}\rightarrow 0$. In the large entropy limit, for the A-type the relevant light tower is that of the winding modes, while for the B-type the relevant one is that of KK modes. On the other hand, in the small entropy limit the situation is reserved. All this finds nice agreement with the results of \cite{Bonnefoy:2019nzv}, which investigated the limit of large entropy of $\mathcal{N}=2$ extremal black holes, for fixed magnetic charges.

It is however important to emphasise that there also exists other types of black holes, with different combinations of electric and magnetic charges being kept fixed. They behave in a different way and do not necessarily lead to massless towers of KK or winding modes for large or small entropy. These particular examples are black holes with charges such that the volume in \eqref{Vextremalp} is kept fixed (see Section (\ref{rescalingcharges}) for more details). Then, the limits ${\cal S}\rightarrow0$ or $\mathcal{S} \to\infty$ do not lead to a light tower of states.

\subsubsection{Non-extremal black holes}

We consider now thermodynamic limits of non-extremal black holes. Before discussing the A- and B-type, a somehow special case of non-extremal black hole is given by the Schwarzschild solution, where all electric and magnetic charges are turned off. Then, one is precisely moving along the Schwarzschild hyperbola ${\cal S}=1/(16\pi{\cal T}^2)$, which is the boundary of the colored region in Figure \ref{fig:BHpopulation}. This is expected, since for a neutral black hole in ${\cal N}=2$ supergravity the scalar fields are constant and independent of ${\cal S}$ and ${\cal T}$. Hence, the volume of the internal space is also constant and equal to its value at infinity, $ \mathcal{V}= \mathcal{V}_\infty = ( 2\sqrt{2} L_\infty^0)^{-2}$. In the limits of large or small asymptotic values, $L_\infty^0\to0$ or $L_\infty^0\to\infty$, one can get light KK or winding modes and the distance conjecture with respect to the internal moduli fields applies.

We now move to black holes with non-vanishing electric and magnetic charges and focus in particular on the type A and B introduced in the previous section. As before, other cases of combinations of charges to be hold fixed are possible.

\subsubsection*{A-type black holes}

At the horizon, $\rho\to-\infty$, the formula \eqref{volumeCY3}  of the internal volume reduces to
\begin{equation}\boxed{
\mathcal{V}_h= \frac{a_0^2}{\sqrt{a_0 a^1 a^2 a^3}} = 64 \pi a_0^2 {\cal S} {\cal T}^2.}
\end{equation}
The dependence of the volume in terms of ${\cal S}$ and $\cal{T}$ is given once we substitute the expression for $a_0$ in eq.~\eqref{a0b0} and use  eq.~\eqref{nonextremalityrelation}. We obtain
\begin{equation}\label{Vnonextremal}\boxed{
\mathcal{V}_h= \frac{4\pi \mathcal{V}_\infty}{\cal S} \bigg({\cal S}{\cal T}+\sqrt{{\cal S}^2 {\cal T}^2 + \frac{2q_0^2}{{\cal V}_\infty}}\bigg)^2\,,}
\end{equation}
which is the expression of the internal volume, when we keep the electric charge $q_0$ constant, while entropy  and temperature  vary together with the black hole mass and the magnetic charges.

First, we consider two different limits for the temperature, for which the internal volume either becomes small or large:
\begin{itemize}
\item In the {\it small temperature} limit (case (II) in Figure \ref{fig:BHpopulationandlimits}), ${\cal T}\rightarrow 0$, we recover eq.~\eqref{Vextremalq} of the extremal case. If we follow a trajectory such as ${\cal S}({\cal T})=1/(8\pi n {\cal T}^2)$, with $n>2$, then we are sure to be within the validity regime of charged black holes, that is below the Schwarzschild hyperbola ${\cal S}=1/(16\pi {\cal T}^2)$. In the small temperature limit, along this trajectory we have
\begin{equation}
{\cal T}\rightarrow0\, :\qquad \mathcal{V}_h\simeq{q_0^2{\cal T}^2}\rightarrow0\, ,\quad m_{KK}\simeq (q_0{\cal T})^{-1/3}\rightarrow\infty\, .\label{smallT}
\end{equation}
The T-dual winding modes behave in an opposite way:
\begin{equation}
{\cal T}\rightarrow0\, :\qquad  m_{wind}\simeq (q_0{\cal T})^{1/3}\rightarrow0\, .\label{smallTwind}
\end{equation}

\item The {\it large temperature} limit (case (I) in Figure \ref{fig:BHpopulationandlimits}), ${\cal T}\rightarrow \infty$, is somewhat more delicate to treat. In this limit, the relevant contributions to the volume are
\begin{equation} \label{VTinf}
\mathcal{V}_h =  16 \pi \mathcal{V}_\infty {\cal S} {\cal T}^2+\frac{4 \pi  q_0^2}{\cal S}\, .
\end{equation}
It is important to note that this limit necessarily corresponds to small entropy, ${\cal S}\rightarrow 0$, for a charged black hole, since the maximum entropy for a given temperature is bounded from above by the hyperbola ${\cal S}=1/(16\pi{\cal T} ^{2})$, corresponding to a Schwarzschild black hole (see Figure \ref{fig:BHpopulation}). This implies that the first term of eq.~\eqref{VTinf} approaches a constant, namely the asymptotic value of the volume at infinity, while the second term diverges. We can follow again a trajectory below the limiting hyperbola, such as ${\cal S}({\cal T})=1/(8\pi n {\cal T}^2)$.
Then, the large temperature limit provides
\begin{equation}
{\cal T}\rightarrow\infty\, :\quad \mathcal{V}_h\simeq{q_0^2{\cal T}^2}\rightarrow\infty\, ,\quad m_{KK}\simeq (q_0{\cal T})^{-1/3}\rightarrow0\, ,
\quad  m_{wind}\simeq (q_0{\cal T})^{1/3}\rightarrow\infty\, .\label{largeT}
\end{equation}
We see that the limit of large temperature always corresponds to decompactification with $\mathcal{V}_h \rightarrow \infty$. The KK masses will then be vanishing, thus leading to the breakdown of the effective field theory. 

\end{itemize}

\vskip0.3cm
Then, we consider similar limits for the entropy:
\begin{itemize}
\item For {\it small entropy} (case (I) in Figure \ref{fig:BHpopulationandlimits}), we recover the extremal limit with eq.~\eqref{Vextremalq} for the volume $\mathcal{V}_h$, similarly to the small temperature limit. However, in this case the volume will become large scaling as
\begin{equation}
 \mathcal{V}_h\simeq 2\pi\frac{q_0^2}{\cal S} \rightarrow \infty\,.
\end{equation}
 Then, for constant electric charge, the limit of small entropy necessarily corresponds to a light tower of KK modes, with masses $m_{KK}\rightarrow 0$.

\item In limit of {\it large entropy} (case (II) in Figure \ref{fig:BHpopulationandlimits}), the volume at leading order is
\begin{equation}
\mathcal{V}_h = 16 \pi \mathcal{V}_\infty {\cal S} {\cal T}^2\,,
\end{equation}
which matches the first term of eq.~\eqref{VTinf}.  We note that the limit ${\cal S}\rightarrow \infty$ necessarily corresponds to the limit of small temperature, ${\cal T}\rightarrow 0$, since both values are upper bounded by the Schwarzschild hyperbola ${\cal T}\sim {\cal S}^{-1/2}$ . This is analogous and parallel to the situation we had for the limit of large temperature. Fixing again a trajectory ${\cal S}={\cal S}({\cal T})$ below the limiting Schwarzschild hyperbola, the volume will become small in the limit of large entropy. The relevant light tower invalidating the effective description is represented thus by winding modes.
\end{itemize}

\subsubsection*{B-type black holes}

\noindent
At the horizon, $\rho\to-\infty$, we can write the internal volume as 
\begin{equation}\boxed{
\mathcal{V}_h= \frac{1}{(64 \pi)^3 (a^1 a^2 a^3)^2 {\cal S}^3 {\cal T}^6}\,.}
\end{equation}
We can obtain the dependence of the volume entirely in terms of the entropy ${\cal S}$ and the temperature ${\cal T}$ by using eq.~\eqref{a1b1} and eq.~\eqref{nonextremalityrelation}. We have
\begin{equation}\label{Vnonextremalp}\boxed{
\mathcal{V}_h= \frac{ 8(M^1_\infty M^2_\infty M^3_\infty)^2 {\cal S}^3}{\pi ^3 \prod_{i=1}^3\left(\sqrt{4(M^i_\infty)^2 (p^i)^2+ {\cal S}^2 {\cal T}^2}+ {\cal S} {\cal T}\right)^2}\,,}
\end{equation}
which is the expression of the internal volume at \emph{fixed magnetic charges}.

\vskip0.2cm

\noindent
Now, we can again consider the following limits for the temperature:
\begin{itemize}
\item In the limit  ${\cal T} \rightarrow 0$ (case (II) in Figure \ref{fig:BHpopulationandlimits}) the volume at leading order  precisely agrees with \eqref{Vextremalp}. We can follow a trajectory ${\cal S}\approx 1/{\cal T}^2$, which stays below the Schwarzschild hyperbola, and obtain
\begin{equation}
{\cal T}\rightarrow0\, :\qquad \mathcal{V}_h  \simeq          \frac{1}{(2\pi)^3}\frac{1} {{\cal T}^6(p^1 p^2 p^3)^2}    \rightarrow\infty\, ,\quad m_{KK}\rightarrow0\, ,\quad m_{wind}\rightarrow\infty\, .\label{smallTm}
\end{equation}
\item The limit of large temperature ${\cal T} \rightarrow \infty$ (case (I) in Figure \ref{fig:BHpopulationandlimits})
necessarily implies ${\cal S} \rightarrow 0$. Being close to the hyperbola ${\cal T}=(16\pi{\cal S})^{-1/2}$, we get that the internal volume becomes infinitesimally small, $\mathcal{V}_h\rightarrow 0$, and thus an infinite tower of light winding modes:
\begin{equation}
{\cal T}\rightarrow\infty\, :\qquad \mathcal{V}_h  \simeq          \frac{1}{(2\pi)^3}\frac{1} {{\cal T}^6(p^1 p^2 p^3)^2}    \rightarrow0\, ,\quad m_{KK}\rightarrow\infty\, ,\quad m_{wind}\rightarrow0\, .\label{largeTm}
\end{equation}
\end{itemize}

The analogous limits for the entropy are:

\begin{itemize}
\item  For {\it small entropy} (case (I) in Figure \ref{fig:BHpopulationandlimits}),   the volume goes to zero:
\begin{equation}
\mathcal{S}\to 0:, \qquad \mathcal{V}_h\propto  \frac{1}{(2\pi)^3}\frac{{\cal S}^3} {(p^1 p^2 p^3)^2}
 \rightarrow 0\, ,
\end{equation}
thus leading to a light tower of winding modes with masses approaching zero.

\item For {\it large entropy} (case (II) in Figure \ref{fig:BHpopulationandlimits}), ${\cal S} \rightarrow \infty$, the volume has an identical expansion as in the limit of large temperature considered above. After fixing a proper trajectory in the ${\cal T}-{\cal S}$ plane, we obtain that the limit of large entropy corresponds to large volume $\mathcal{V}_h\rightarrow \infty$ and a light tower of KK modes. This result is somehow analogous to the case of extremal black-holes, with fixed magnetic charges, seen in Sec.~\ref{Sec.Extremal} and investigated in \cite{Bonnefoy:2019nzv}.

\end{itemize}

In summary, we see that the limit (I) in Figure \ref{fig:BHpopulationandlimits} (for the chosen combinations of charges) leads either to a tower of light KK particles  or a tower  of light winding modes, measured at the horizon of the black hole. This is the limit of small black holes with large temperatures, which behave like particles states. Due to the presence of infinite light modes, in accordance with the SDC we conclude that the effective field theory on the horizon breaks down. A similar conclusion can be drawn for the limit (II) of large black holes with large entropy and small temperature.

It is interesting to notice that the volume of the internal manifold ${\cal V}$ also appears in the supergravity Lagrangian \eqref{S4dSG} as the gauge coupling of the U(1) gauge fields in the matrix $\mathcal{I}_{\Lambda\Sigma}$, as one can check from \eqref{symm-matr-N}. One finds in fact that the $F^0_{\mu\nu}$ gauge field has coupling equal the internal volume ${\cal I}_{00}=-{\cal V}$. From the analysis in this section we see that near the horizon, in addition to the KK masses and winding modes, the limits discussed above also affect the U(1) couplings. This observation connects our results to yet another swampland conjecture, namely the absence of global symmetries in quantum gravity.

\subsubsection{Rescaling of all charges}\label{rescalingcharges}

As we already mentioned before, the light towers of states appear in the limit of large/small temperature and entropy in case certain combinations of electric or magnetic charges are kept fixed. However, if all charges are increased or decreased in the same way, the internal volume stays constant. Below, we investigate this particular scaling limit in some detail.

Let us consider a black hole of the ${\it STU}$ model with charges $q_0,p^1,p^2,p^3$, and let us study the transformation that rescales all charges in the same way.  We start from the extremal case, where the metric is
\begin{align}\label{metric-extremal-tau}
ds^2=- e^{2U} dt^2 + e^{-2U}\left(\frac{d\rho^2}{\rho^4} + \frac{1}{\rho^2}d\Omega_{2}^2\right)\ .
\end{align}
A rescaling of all charges $q_0\to \lambda q_0$ and $p^i\to \lambda p^i$, together with the coordinates rescaling $\rho\to \lambda^{-1}\rho$, leaves the functions $I_0,I^i$ invariant, thus also $e^{2U}\to e^{2U}$, and $z^i(r)\to z^i(r)$. If we also redefine the time coordinate by $t \to \lambda t$, the metric becomes
\begin{align}
ds^2 \to \lambda^2 \left( - e^{2U} dt^2 + \left(e^{-2U}\frac{d\rho^2}{\rho^4} + \frac{1}{\rho^2}d\Omega_{2}^2\right)\right),
\end{align}
namely the charge rescaling and coordinate redefinition result in a Weyl rescaling of the metric. 

The same also happens for the non-extremal solution. Considering the family of rescalings
 \begin{align}
 \label{eq:rescaling-nonex}
 q_0 \to \lambda q_0 \ , \qquad   p^i \to \lambda p^i \ , \qquad  \rho \to  \lambda^{-1} \rho \ , \qquad    c \to \lambda c \ , \qquad t\to \lambda t \ ,
    \end{align}
the quantities $a_0$, $a^i$ and $b_0$, $b^i$ are left invariant, so again  $e^{2U}\to e^{2U}$, and $z^i(\rho)\to z^i(\rho)$, while
\begin{align}
\mathcal{S} \to \lambda^2 \mathcal{S} \ , \qquad\mathcal{T} \to \lambda^{-1}\mathcal{T}  \ .
\end{align}
On the metric \eqref{metric}, this corresponds to
\begin{align}
  ds^2 \to \lambda^2 \left(   - e^{2U}dt^2+e^{-2U}  \left[ \frac{c^4 d\rho^2}{\sinh^4(c\rho)}     + \frac{c^2}{\sinh^2(c\rho)} \left(d\theta^2+\sin^2\theta d\phi^2\right) \right]    \right),
    \end{align}
which is again a Weyl rescaling. Now notice that, since the scalars are left invariant, the volume is also invariant \cite{Bonnefoy:2019nzv}.  If one considers this Weyl rescaling within a ten-dimensional embedding of the four-dimensional black hole geometry, then the transformation \eqref{eq:rescaling-nonex} can be interpreted as a rescaling of the Planck Mass.

Notice that the black hole ADM mass, by simple background subtraction at infinity, is
\begin{align}
 M= \frac14 \left(\sqrt{c^2+(4q_0L^0_\infty)^2}+\sum_{i=1}^3 \sqrt{c^2+(4p^iM_{i\infty})^2} \right)\ ,
 \end{align}
which also gets rescaled as $M\to \lambda M$, to preserve the BPS bound of the theory $M\geq |Z|$ (the central charge scales together with the charges, $Z \to \lambda Z$).

\subsection{Thermodynamic dualities}

In this section, we would like to investigate further the existence of new temperature and entropy dualities,  ${\cal T}\leftrightarrow 1/{\cal T}$  or  ${\cal S}\leftrightarrow 1/{\cal S}$, acting as geometric duality on the compact internal space, namely exchanging KK modes with winding modes. We will point out situations in which large and small temperature limits are precisely dual to each other.

Let us pick a fixed symplectic frame for the charges, and let us investigate possible dualities between the large and the small temperature regime of supergravity black holes. In particular, for the {\it STU} model, we consider black holes with four non-vanishing charges $q_0,p^1,p^2,p^3$.

We start from the temperature. Consider A-type black holes and compare the high temperature limit \eqref{largeT} with the small temperature limit \eqref{smallT}. It is easy to see that the transformation 
\begin{equation}
{\cal T}\,\longleftrightarrow\, {1\over q_0^2{\cal T}}\, ,\qquad \qquad \quad\qquad q_0\longrightarrow q_0\ ,
\end{equation}
leaving the electric charge $q_0$ fixed, inverts the internal volume, ${\cal V}_h\longleftrightarrow{1\over{\cal V}_h}$ and hence exchanges KK with winding modes. This is a realization of a ${\cal T}\leftrightarrow 1/{\cal T}$  \textit{temperature duality}.\\
\noindent Then, consider B-type black holes. We see immediately that there is again a {\it high-low temperature duality}, inducing ${\cal V}_h\longleftrightarrow{1\over{\cal V}_h}$ and exchanging the internal KK and winding modes among each other. This has the following form:
\begin{equation}
{\cal T}\,\longleftrightarrow\, {1\over (p^1 p^2 p^3)^{2/3}{\cal T}}\, ,\qquad\qquad p^i\longrightarrow p^i\, ,
\end{equation}
where the magnetic charges $p^i$ are kept fixed.

Similarly to what done for the temperature, we can also point out an entropy duality. For the A-type black holes this is of the form
\begin{equation}
{\cal S}\,\longleftrightarrow\,{q_0^{4}\over{\cal S}}\, ,\qquad\qquad\qquad \qquad q_0\longrightarrow q_0\, ,
\end{equation}
while for B-type black holes we have
\begin{equation}
{\cal S}\,\longleftrightarrow\,{(p^1 p^2 p^3)^{4/3}\over{\cal S}}\, ,\qquad \qquad p^i\longrightarrow p^i\, .
\end{equation}

Note that there is yet another class of entropy/temperature transformations, which however keep the volume ${\cal V}_h$ invariant, but rescale the electric charge $q_0$ (case A) or the magnetic charges $p^i$ (case B) in a non-trivial way:
\begin{equation}
\sqrt{\cal S}\,\longleftrightarrow\,{\cal T} \quad{\rm with}\quad q_0\,\longrightarrow\,{{\cal T}\over\sqrt {\cal S}}~q_0\,\, \,\,(A)
\quad {\rm or} \quad p^i\,\longrightarrow\,{{\cal T}\over\sqrt {\cal S}}~p^i \,\,\,\,(B)\, .
\label{STex1}
\end{equation}
These transformations act like in the Schwarzschild case (see eq.(\ref{STex})) and exchange small with large black holes. Moving on a Schwarzschild like hyperbola, these transformations look like
\begin{equation}
{\cal S}\,\longleftrightarrow\,{1\over {\cal S}}  \quad{\rm and}\quad{\cal T}\,\longleftrightarrow\,{1\over {\cal T}}       \quad{\rm with}\quad   q_0\,\longrightarrow\,{ {\cal T}^2}~q_0  \,\,\,\, (A)
  \quad {\rm or} \quad p^i\,\longrightarrow\,{ {\cal T}^2}~p^i \,\,\,\,(B)   \, .
\label{STdual2}
\end{equation}

\subsection{Embedding of Einstein-Maxwell-dilaton in ${\cal N}$=2 Supergravity}
\label{EMdSUGRA}

The Einstein-Maxwell-dilaton (EMd) black holes of Section \ref{sec:EMDblackholes}, for specific values of parameter $\lambda$, are captured by the $p^0,q_0$ configuration of ${\cal N}=2$ matter coupled supergravity \cite{Ferrara:1996dd, Ceresole:2007rq}, and correspond to the KK black hole in the reduction from five to four dimensions\cite{Gibbons:1985ac,Horowitz:2011cq}. To discuss the supergravity embedding of the KK black hole we consider the ${\it STU}$ model introduced in Section \ref{SG-blackholes}; we set $z^1=z^2=z^3=- i e^{- \sqrt{2/3}\phi}$ and only take the vector field strength $F^0_{\mu\nu}$ to be nonzero \cite{Duff:1999gh}. One can see then that the Lagrangian of this model is of the EMd form
 \begin{align}
\mathcal{I}_{\Lambda\Sigma}  F^{\Lambda}_{\mu\nu} F^{\Sigma\, \mu\nu} \to  \mathcal{I}_{00} (F^{0}_{\mu\nu})^2 &= e^{-\sqrt6 \phi} (F^0_{\mu\nu})^2 \ ,\\  
g_{i\bar\jmath}\partial_\mu z^i\partial^\mu \bar z^{\bar \jmath} &\to \frac12 \partial_\mu \phi \partial^\mu \phi\ ,
\end{align}
for a value $\lambda = \sqrt{3/2}$.

In the ${\it STU}$ model, the properties of extremal black holes are captured by the U-duality group quartic invariant \cite{Bellucci:2008sv}
\begin{align}
\label{quartic-inv}
\mathcal{J}_4 = - (p^0q_0+p^i q_i)^2 +4 \sum_{i<j} p^iq_i p^j q_j-4 p^0 q_1 q_2 q_3 + 4 q_0 p^ 1 p^ 2 p^ 3\ .
\end{align}
One sees that for a choice of charges $p^0,q_0$, this invariant is negative definite, thus it corresponds to a configuration which does not preserve supersymmetry in the extremal limit (non-BPS) (see \cite{Bellucci:2008sv} and references therein). For such choice of charges, the extremal black hole we obtain is given by (see \cite{Ceresole:2009jc,Ferrara:2006em} and references therein)
\begin{align}
z^1 z^2 z^3|_{h} &=  i \frac{q_0}{p^0} \ ,  \\
e^{-\sqrt{2/3}\phi}|_{h} &= \left( \frac{q_0}{p^0}  \right)^{1/3}  \ ,   \\
\frac{{\cal S}}{4\pi}&= \sqrt{\mathcal{J}_4(p^0,q^0)} =|p^0 q_0| \ ,
\end{align}
which has a residual $\left(SO(1,1)\right)^2$ moduli space \cite{Bellucci:2008sv}. This reproduces the extremal black hole behavior of the EMd models.

The complete flow from asymptotic infinity to the horizon is given by the extremal metric \eqref{metric-extremal-tau} with 
 \begin{align}
e^{-2U(\rho)}&=\sqrt{I_0(\rho) \tilde I_0(\rho)}. \\ 
 z^i &= - i \left(\frac{q_0}{p^0} \right) ^{1/3} \sqrt{\frac{\tilde I_0(\rho)}{I_0(\rho)}}, \\
 I_0 &= \left(a-\sqrt{q_0 p^0}\rho\right)^2-b \ ,  \\
 \tilde I_0 &=  \left(a-\sqrt{p^0 q_0}\rho\right)^2+b \ ,
  \end{align}
where $a\neq0$ and $b$ are real constants.
Notice that this is a case where the extremal solution has running scalars, since the values of the scalars vary throughout the radial direction. This black hole is the static extremal limit of the rotating black hole constructed in \cite{Bena:2009ev}.

\subsection{Small black holes}

The configuration with only one of $p^0$ or $q_0$ turned on corresponds to \emph{small black hole} solutions of ${\cal N}=2$ matter coupled supergravity. However, there are also more general small black hole configurations in ${\cal N}=2$ supergravity.

Small black holes are a class of extremal black hole solutions of ${\cal N}=2$ supergravity arising for a choice of charges such that the quartic U-duality invariant of the theory (given in \eqref{quartic-inv}) is ${\cal J}_4=0$. They are relevant for our discussion since first the scalar fields for this solutions never reach a fixed point at finite distance in moduli space\footnote{Contrary to the attractor mechanism for finite area extremal black holes in ${\cal N}=2$ supergravity.} \cite{Ceresole:2010nm}, and second they arise as a $I_4\to 0$ limit of finite-area black holes, for an appropriate tuning of the charges. Thus, they appear as solutions naturally associated to a notion of infinite distance in moduli space. 

In the case of the $STU$ model, also the configurations $p^0, q_1,q_2,q_3$ and $q_0,p^1,p^2,p^3$ become small black holes when $p^0 \to 0$ and $q_0 \to 0$ respectively, as can be seen from \eqref{quartic-inv}. To illustrate what happens let us consider the limit $q_0\to 0$ in the non-extremal solution considered for the $STU$ model in Section \ref{sec:non-extremal} (which corresponds to setting $b_0=0$ in the simpler solution \eqref{a0b0}). Entropy and temperature are
\begin{align}
\mathcal{S}|_{q_0=0} \ & \sim \sqrt{ c} \  \prod_{i=1}^3\sqrt{c+\sqrt{c^2+\left(4 p^iM_{i\infty}\right)^2}} \ , \\
\mathcal{T}|_{q_0=0} \ & \sim \frac{\sqrt c}{   \prod_{i=1}^3\sqrt{c+\sqrt{c^2+\left(4p^iM_{i\infty}\right)^2}} \ ,}
\end{align}
while scalars run from their value at infinity $z^i=-i\lambda_\infty^i$ to the horizon value:
 \begin{align}
  z^i_{h}=-i  \, \sqrt{2c}\, \lambda_\infty^1  \left( \frac {c+\sqrt{c^2+(4 p^1 M_{1\infty})^2}} { \left( c+\sqrt{c^2+ (4 p^2 M_{2\infty})^2} \right)  \left(  c+\sqrt{c^2+ (4 p^3 M_{3\infty})^2} \right) } \right)^\frac12 \ .
  \end{align}
Thus, we see that in the extremal limit $c\to 0$ (corresponding again to ${\cal T}\to0$)
\begin{align}
 \mathcal{S}|_{q_0=0} &\to 0 \ , 
 \end{align}
and the scalars run from a constant value at infinity to $z_h^i = 0$ at the horizon. In the simpler solution of the previous section, this limit corresponds to 
\begin{align}
 e^{-\sqrt{2/3}\phi}\to 0 \,, \qquad \phi \to \infty \
 \end{align}
and it is again a limit at infinite distance in moduli space. Thus, we can extend to this setup the conclusion that small black holes are outside of the regime of validity of the supergravity effective theory, as we discussed for the EMd case in the previous section. 

\subsubsection{Doubly small black holes}

There is another interesting configuration that one can analyse, starting from the $STU$ black hole with charges $q_0,p^1,p^2,p^3$ (or $p^0,q_1,q_2,q_3$, analogously). Let us consider again the non-extremal solution of Section \ref{sec:non-extremal} and now turn off both $q_0$ and one of the magnetic charges, say $p^1$. In this case, entropy and temperature read
\begin{align}
\mathcal{S}|_{q_0=0,p^1=0}\ \  & \sim \ \ c \left(\left(c+\sqrt{c^2+(4{p}^2 M_{2\infty})^2}\right)\left(c+\sqrt{c^2+(4{p}^3M_{3\infty})^2}\right)\right)^\frac12 \ ,  \\
\mathcal{T}|_{q_0=0,p^1=0} \ \ & \sim \ \ \frac{1}{\left(\left(c+\sqrt{c^2+(4{p}^2 M_{2\infty})^2}\right)\left(c+\sqrt{c^2+(4{p}^3M_{3\infty})^2}\right)\right)^\frac12}\ ,
\end{align}
while the horizon value of the scalars is  
\begin{align}
 z^i_{h}&=-i \, \lambda_\infty^i \frac{2c}{\left(\left(c+\sqrt{c^2+(4{p}^2 M_{2\infty})^2}\right)\left(c+\sqrt{c^2+(4{p}^3M_{3\infty})^2}\right)\right)^\frac12} \ .
\end{align}
Again, these are regular solutions at finite temperature but, as $c\to0$, they have vanishing entropy and finite temperature, and the scalars at the horizon $z_{h}^i\to 0$. This limit is once more at infinite distance in moduli space. These solutions are thus supergravity examples of the extremal limit $t_1$ illustrated in Figure \ref{fig:BHpopulationandlimits}. Let us stress once more that this limit corresponds to a breakdown of the effective description associated with a light tower of states predicted by the SDC.

\section{Conclusion}
\label{sec:concl}

In this work, we have studied thermodynamic properties of non-extremal charged black holes in relation to the Swampland program. The ${\cal T}-{\cal S}$ phase diagram, in terms of the temperature and entropy, has been a main focus of our investigation. Specifically, the analysis of asymptotic regions, as shown in Figure \ref{fig:BHpopulationandlimits}, has turned out to provide novel insight into the relation between the Swampland Distance Conjecture and the physics of black holes.

While an application of the metric distance in these non-compact backgrounds turns out to be problematic in the case of finite entropy extremal black holes, the field distance traversed by the moduli fields from spatial infinity to the black hole horizon successfully identifies the problematic regions. This was also confirmed by embedding the effective descriptions, Einstein-Maxwell-dilaton theory as well as $\mathcal{N} = 2$ supergravity, in string theory and identifying the light towers suggested by the Swampland Distance Conjecture in the vicinity of the black hole.

\begin{figure}[h]
\begin{center}
\begin{tikzpicture}[scale=1.25]
\draw[->, line width=.5mm] (0, 0) -- (4.2, 0) node[below] {${\cal X}$};
\draw[->,line width=.5mm] (0, 0) -- (0,4.33) node[left] {${\cal Y}$};
\draw[scale=1.5,domain=0.:2.,smooth, variable=\x,blue, thick,line width=.6mm] plot ({\x},{-(\x)+2)});
\fill [blue!25, domain=0.:2., variable=\x]  (0.0, 4.15) -- plot[scale=1.5,domain=0.:2.] ({\x},{-(\x)+2)})
  -- (0.02, 0.02);
\draw[red, line width=.7mm] (0, 0) -- (3, 0);
\node () at (3,-.3) {$\frac \pi2$}; 
\node () at (-.3,3) {$\frac \pi2$}; 
\node () at (3,3) {excluded region};
\draw[red,fill=red] (0,3) circle[radius=2pt];
\draw[red,fill=red] (3,0) circle[radius=2pt];
\end{tikzpicture}
\end{center}
\caption{Version of the diagram in Figure \ref{fig:BHpopulation} with the point at infinity included (red dots). The coordinates are related as $\mathcal{Y} =\arctan \mathcal{S}$ and $\mathcal{X} = \arctan (2 \mathcal{T}^2)$. The red points and line are regions where towers of infinitely-many massless states are expected to emerge. The blue line $\mathcal{Y} = - \mathcal{X} + \frac{\pi}{2}$ is the Schwarzschild hyperbola $\mathcal{S} = (2\mathcal{T}^2)^{-1}$ in the new coordinates.}
\label{fig:BHpopulationcompact}
\end{figure}
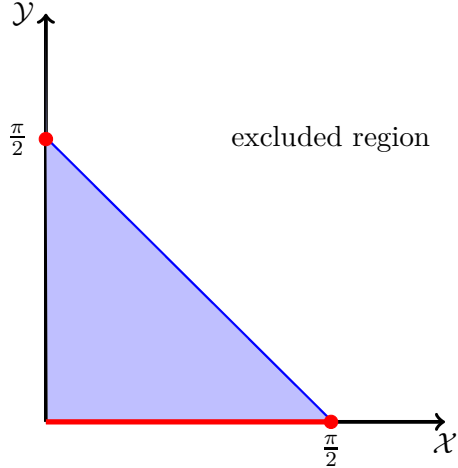

Our findings are visually summarized in Figure \ref{fig:BHpopulationcompact}, which proposes again the temperature-entropy phase diagram of non-extremal black holes, but now in a ``compactified" form. The points at infinity are shrunk to $\frac{\pi}{2}$ from the origin and the validity region for RN black holes becomes a finite triangle. The loci where massless towers of states are expected to emerge are highlighted in red.

The large/small entropy limit, which induces the infinite field distance, is also closely related to the AdS distance conjecture \cite{Lust:2019zwm} and the scale separation conjecture \cite{Tsimpis:2012tu,Gautason:2015tig}, which state that in the limit of small AdS curvature a massless tower of states should emerge. For four-dimensional extremal black holes, the near horizon geometry has the form $AdS_2\times S^2\times M_6$, where $M_6$ is an internal space of volume ${\cal V}_h$. The problem of scale separation for these kind of $AdS_2$ geometries was recently investigated in \cite{Lust:2020npd}. In the large entropy limit, ${\cal S}\rightarrow\infty$, the radii of $AdS_2$ and $S^2$ become large and according to the AdS distance conjecture a massless tower of states would be expected.\footnote{Again, we want to advise caution in the direct application of Swampland constraints in two and three dimensions.} This tower can be identified with the KK modes of the $S^2$. In this paper, we are rather investigating the question of scale separation between the four-dimensional black hole and the internal space $M_6$. 
As we have seen, this scale separation is possible if all electric and magnetic charges scale in the same way (since the volume remains constant while the entropy can vary - see Sec.~\ref{rescalingcharges}), but it is not possible if either the electric or the magnetic charges are kept fixed. For non-extremal black holes, the near horizon geometry is instead ${\cal R}^{1,1}\times S^2\times M_6$.  The scale of the 2-dimensional Rindler space ${\cal R}^{1,1}$ is determined by the surface gravity $\kappa=2\pi{\cal T}$. Then, the question of scale separation and the emergence of massless tower of states can be rephrased in terms of a {\it temperature distance} and one can study the limits ${\cal T} \rightarrow \{0,\infty\}$. 

Another intriguing question addressed in our paper regards the existence of thermodynamic dualities between small and large temperature as well as entropy, i.e. between small and large black holes. Our motivation comes from the fact that temperature dualities are present in several statistical models and that black holes allow in fact for a statistical interpretation.  Already at the level of a Schwarzschild black holes, we have observed the existence of such dualities (expressed in terms of $\cal T$ and $\cal S$), which exchange large and small black holes, thus acting on the mass as $M \leftrightarrow 1/M$. 
More interestingly, in the context of non-extremal black holes in ${\cal N}=2$ supergravity, we have shown that temperature- and entropy-dualities act on the internal volume $\cal V$ by exchanging light KK and winding states.
With this observation at hand, we do not claim that large and small black holes are physically equivalent to each other but, when coupled to scalar field of the effective field theory, they can correspond to an equivalent internal (dual) spectrum.  We interpret such thermodynamic dualities as the effective remnants of the string dualities (such as T- or S-dualities). 
Clearly, also the existence of such dualities, like the exchange of small and large entropy (or small and large mass), implies having control over higher derivative corrections, which mainly come into play for small ${\cal S}$, that is when the supergravity regime breaks down. We leave the question whether our findings are stable against such corrections for future work.

Finally, let us mention a few more questions and possible generalizations of the present work. 
First, as already mentioned, the limit of small entropy ${\cal S}\rightarrow 0$ is very subtle, requires very small electric or magnetic charges, and most likely needs some refinement of the effective two-derivative supergravity action.
Here we expect that higher derivative corrections to the effective action 
could change the relation between the thermodynamic black hole properties and the internal moduli fields.
Second, the discussion could be extended to rotating non-extremal solutions, using the stationary solutions in ${\cal N}=2$ of \cite{Behrndt:1997ny} or the non-BPS black rings, which are considered in \cite{Bena:2007kg,Bena:2009ev}. Another interesting problem is to investigate the similarity of black hole potentials and the scalar field flux potentials in extended supergravity. In this context, there might arise some interesting analogies between the formalism of the black hole limits considered in this paper and of the limits for supersymmetry breaking in the context of the gravitino conjecture \cite{Cribiori:2021gbf,Castellano:2021yye}.

\vskip0.5cm
\vspace{10px}
{\bf Acknowledgements}
\vskip0.1cm
\noindent

We thank Mirjam Cveti\v{c} and Gary Shiu
for useful discussions. 
The work of N.C.~is supported by the Alexander von Humboldt foundation.
The work of D.L.~is supported  by the "Origins" Excellence Cluster and also by the German-Israelian-Project (DIP) "Holography and the Swampland". 

\appendix

\section{$\mathcal{N}=2$ four-dimensional supergravity and BPS black holes: notation and conventions \label{appendix-supergravity} }

In this appendix, we summarize notation and conventions used in Section \ref{SG-blackholes}. For a comprehensive review on $\mathcal{N}=2$ special geometry and BPS black holes in supergravity we refer to \cite{Andrianopoli:1996cm} and \cite{Andrianopoli:2006ub}.

The bosonic sector of four-dimensional $\mathcal{N}=2$ supergravity coupled to $n_V$ vector multiplets and $n_H$ hypermultiplets includes the Einstein-Hilbert term, the special K\"ahler nonlinear sigma model of $n_V$ complex scalars $z^i$ coming from the vector multiplets (with scalar manifold $\mathcal{M}_{V}$), $(n_V+1)$ gauge fields $F^\Lambda$ with electric and magnetic couplings to the $z^i$, and a quaternionic non-linear sigma model of $4 n_H$ real scalars $q^u$ from the hypermultiplets (with scalar manifold $\mathcal{M}_H$). In the ungauged case, hypermultiplets can be decoupled from the other bosonic fields, therefore we are left with 
\begin{align}
\label{S4dSG}
S= \int d^4x \sqrt{\textrm{-} g} \left(-\frac{R}{2}+ g_{i\bar\jmath}\partial_\mu z^i\partial^\mu \bar z^{\bar\jmath}  + \mathcal{I}_{\Lambda\Sigma}F^{\Lambda}_{\mu\nu}F^{\Sigma\,\mu\nu} + \frac{1}{2\sqrt{\textrm{-} g}}\mathcal{R}_{\Lambda\Sigma} \, \epsilon^{\mu\nu\rho\sigma}F^\Lambda_{\mu\nu}F^\Sigma_{\rho\sigma}\right) \ .
\end{align}
The special K\"ahler metric $g_{i\bar \jmath}$, together with the couplings $\mathcal{I}_{\Lambda\Sigma}(z^i \,,\bar z^{\bar\imath})$ and $\mathcal{R}_{\Lambda\Sigma}(z^i\,, \bar z^{\bar\imath})$, are determined by the symplectic \emph{prepotential} $F(X^\Lambda)$, which specifies the supergravity model under consideration. 

The {\it STU} model investigated in the main text is encoded into the prepotential
\begin{equation}
\label{stu-prepot-cub}
F(X^\Lambda)=\frac{ X^1 X^2 X^3}{X^0} \ .
\end{equation}
The three complex scalars $z^i=(S,T, U)$, are related to the projective holomorphic coordinates $X^\Lambda(z^i)$ as 
\begin{equation}
\label{proj-coord}
z^i=\frac{X^\Lambda}{X^0}\,,
\end{equation}
which in turn can be organised into holomorphic symplectic sections 
\begin{equation}
\label{Omega}
\Omega = \begin{pmatrix}
X^\Lambda  (z^i)\\ F_\Lambda(z^i) \end{pmatrix} \ ,
\end{equation}
where $F_\Lambda= \frac{\partial}{\partial X^\Lambda} F(X^\Sigma)$.
The K\"ahler potential can be written as a symplectic product 
\begin{equation}
\mathcal{K}(z^i\,,\bar z^{\bar \imath})= - \log \left[i (\overline{X}^\Lambda F_\Lambda- \bar F_\Lambda X^\Lambda)  \right] \equiv -\log\left(i\, \langle \overline{\Omega}\,, \Omega\rangle\right)\ ,\end{equation}
giving the metric $g_{i\bar\jmath}=\partial_i \bar\partial_{\bar\jmath}\mathcal{K}$. The expression inside the logarithm is proportional to the volume of the Calabi-Yau manifold of the associated string compactification.

One can also define covariantly holomorphic symplectic sections as 
\begin{equation}
V (z^i\,,\bar z^{\bar \imath})= e^{\mathcal{K}/2}\Omega = \begin{pmatrix}
e^{\mathcal{K}/2}X^\Lambda  \\ e^{\mathcal{K}/2} F_\Lambda \end{pmatrix} 
\equiv \begin{pmatrix} L^\Lambda  (z^i\,,\bar z^{\bar \imath})\\ M_\Lambda(z^i\,,\bar z^{\bar \imath}) \end{pmatrix} \ ,
\end{equation}
and, together with their K\"ahler covariant derivatives
\begin{equation}
f^\Lambda_i =\nabla_i L^\Lambda \equiv \left(\partial_i +\frac12 \partial_i\mathcal{K}  \right)L^\Lambda\ , \qquad  h_{\Lambda\,i} =\nabla_i M^\Lambda \equiv \left(\partial_i +\frac12 \partial_i\mathcal{K}  \right)M_\Lambda \ ,
\end{equation}
they can be used to construct the couplings of the scalars to electric and magnetic fields. These are expressed via the matrix $\mathcal{N}_{\Lambda\Sigma}=\mathcal{R}_{\Lambda\Sigma}+i\mathcal{I}_{\Lambda\Sigma}$ as
\begin{equation}
\label{Nlamsig}
\mathcal{N}_{\Lambda\Sigma}=h_{\Lambda\, A}(f^{-1})_{\Sigma}^A \ ,  \qquad\qquad A=0,1...,n_V
\end{equation}
where
\begin{equation}
\label{fhinN}
f^\Lambda_{\ A}=(L^\Lambda,(f^\Lambda_{i})^*)\ , \qquad h_{\Lambda\, A}=(M_\Lambda,(h_{\Lambda\,i})^*)\ .
\end{equation}

One can introduce the \emph{central charge} $Z(z^i,\bar z^{\imath}, q_\Lambda, p^\Lambda)$, which governs the supersymmetric properties of ${\cal N}=2$ BPS black holes \cite{Andrianopoli:2006ub}. It is a symplectic invariant constructed out of the symplectic sections and the electric-magnetic charge vector $\Gamma=(p^\Lambda,q_\Lambda)$ as
\begin{equation}
Z(z^i,\bar z^{\imath}, q_\Lambda, p^\Lambda) \equiv \langle \Gamma\,,{V} \rangle = L^\Lambda(z^i,\bar z^{\imath}) q_\Lambda - M_\Lambda(z^i,\bar z^{\imath}) p^\Lambda \ .
\end{equation}
BPS black holes have a universal near horizon Bertotti-Robinson geometry, proper of extremal Reissner-Nordstrom black holes, namely $AdS_2\times S^2$. Supersymmetry further requires $\ell_{AdS_2}=R_{S^2}$, and these are determined by the central charge value at the horizon, so that the entropy can be expressed as 
\begin{align}
{\cal S}_{BPS}(q_\Lambda,p^\Lambda)&=\pi |Z(z^i_{h},\bar z^{\bar\imath}_{h},q_\Lambda,p^\Lambda)|\ .
\end{align}
However, due to the attractor mechanism the scalar fields at the horizon are completely fixed by the charges through the equations\cite{Ferrara:1995ih,Ferrara:1996dd,Ferrara:1996um}
\begin{align}
\partial_i|Z(z^i,\bar z^{\bar\imath},q_\Lambda,p^\Lambda)|_{h}&=0 \ ,
\end{align}
and thus the entropy is actually a function of the charges only
\begin{align}
{\cal S}_{BPS}(q_\Lambda,p^\Lambda)&=\pi |Z(q_\Lambda,p^\Lambda) |_{h}\ .
\end{align}

For the ${\it STU}$ model, with the projective coordinates choice $X^0=1$, the relevant special geometry quantities are  
\begin{align}
X^\Lambda &= \begin{pmatrix}
1 \\ z^1 \\ z^2 \\ z^3
\end{pmatrix} \ ,\qquad 
F_\Lambda = \begin{pmatrix}
- z^1 z^2 z^3\\ z^2 z^3 \\ z^1 z^3 \\ z^1 z^2 
\end{pmatrix} \ ,
\end{align}
the K\"ahler potential is 
\begin{align}
e^{-\mathcal{K}}= -i(z^1-\bar z^{\bar 1})(z^2-\bar z^{\bar 2})(z^3-\bar z^{\bar 3}) = 8 \mathcal{V}\ ,
\end{align}
($\mathcal{V}$ being the volume of the internal compact manifold) and the metric is 
\begin{align}
g_{i\bar\jmath}= - \frac{1}{(z^i-\bar z^{\bar \jmath})^2} \delta_{i\bar\jmath} \ .
\end{align}

The symplectic matrix $\mathcal N_{\Lambda\Sigma}$ can be obtained then from \eqref{Nlamsig} and \eqref{fhinN}. Setting Re$z^i=0$ and  Im$z^i\equiv \lambda^i$, this matrix becomes 
\begin{align}\label{symm-matr-N}
\mathcal N_{\Lambda\Sigma}&= i \mathcal{I}_{\Lambda\Sigma}=-i\left( \begin{array}{cccc}
\lambda^1 \lambda^2 \lambda^3 & 0& 0 &0\\
0 & \frac{\lambda^2 \lambda^3}{\lambda^1} &0 &0\\
0  & 0 & \frac{\lambda^1 \lambda^3}{\lambda^2} &0\\
0 & 0 & 0 & \frac{\lambda^1 \lambda^2}{\lambda^3}
\end{array} \right) \ .
\end{align}
We notice that $\mathcal{I}_{00}=- \mathcal{V}$, as pointed out in Section \ref{thlimits}.

\subsection{Symplectic duality transformations}

Different functions $F(X)^\Lambda$ can lead to equivalent equations of motion. Such equivalence  involves the electric-magnetic duality of the electro-magnetic field strengths 
\cite{Ceresole:1995jg,deWit:1995dmj,Andrianopoli:1996cm,Andrianopoli:2006ub}.
Specifically,  let us introduce the field strengths tensors $G^{\pm}_{\mu\nu \Lambda}$ as
\begin{equation}
G^+_{\mu\nu \Lambda}={\cal N}_{\Lambda\Sigma}F^{+\Sigma}_{\mu\nu}\,,\quad G^-_{\mu\nu
\Lambda}=\bar{\cal N}_{\Lambda\Sigma}F^{-\Sigma}_{\mu\nu}\, , \label{defG}
\end{equation}
where $F^{\pm}_{\mu\nu} = \frac12 \left(F_{\mu\nu} \pm i \epsilon_{\mu\nu \rho\sigma}F^{\rho\sigma}\right)$. Then, the Bianchi identities and equations of motion for the Abelian gauge fields can be written as
\begin{equation}
\partial^\mu \big(F^{+\Lambda}_{\mu\nu} -F^{-\Lambda}_{\mu\nu}\big)
=0\,,\qquad
\partial^\mu \big(G_{\mu\nu \Lambda}^+ -G^-_{\mu\nu \Lambda}\big) =0\,,
\label{Maxwell}
\end{equation}
which are invariant under the transformations
\begin{align}
F^{+}_{\mu\nu}&\longrightarrow \tilde F^{+\Lambda}_{\mu\nu} =
U^\Lambda{}_\Sigma\, F^{+\Sigma}_{\mu\nu}+Z^{\Lambda\Sigma}\,G^+_{\mu\nu \Sigma}\, ,\\
G^+_{\mu\nu \Lambda}&\longrightarrow \tilde G^+_{\mu\nu \Lambda}= V_\Lambda{}^\Sigma\,
G^+_{\mu\nu \Sigma}+W_{\Lambda \Sigma}\,F^{+\Sigma}_{\mu\nu}\,, \label{FGdual}
\end{align}
where $U$, $V$, $W$ and $Z$ are constant real  $(n_V+1)\times(n_V+1)$ matrices.
The transformations for the anti-selfdual tensors follow by complex conjugation.
To ensure that the rotated fields $\tilde G^\pm_{\mu\nu}$ and $\tilde  F^{\pm}_{\mu\nu}$ are still related by a symmetric matrix $\mathcal{N}$ as in (\ref{defG}), the transformations must be symplectic.
More precisely,
\begin{equation}
{\cal O}\ \mathrel{\mathop{=}\limits^{\rm def}}\
\begin{pmatrix} U & Z \\W & V \end{pmatrix}
\label{uvzwg}
\end{equation}
must be an $Sp(2n_V+2,{\mathbb{R}})$ symplectic matrix, that is it must satisfy
\begin{equation}
{\cal O}^{-1} = \mathcal{J} \, {\cal O}^{\rm T} \,\mathcal{J}^{-1} \qquad
\mbox{where} \quad \mathcal{J} = \begin{pmatrix} 0&1 \\
-1 & 0
\end{pmatrix} \,.
\label{spc}
\end{equation}
For the sub-matrices $U$, $V$, $W$ and $Z$, this means
\begin{eqnarray}
&~&U^{\rm T} V- W^{\rm T} Z = V^{\rm T}U - Z^{\rm T}W =
{\bf 1}\,,\nonumber\cr
&~&U^{\rm T}W = W^{\rm T}U\,,\qquad Z^{\rm T}V= V^{\rm T}Z\,.
\label{spc2}
\end{eqnarray}
It turns out from (\ref{FGdual}) that the kinetic term of the vector fields (\ref{Maxwell}) does not generically preserve its form under $Sp(2n_V+2,{\mathbb R})$, which means that generally it is only the combined equations of motion and Bianchi identities that are symplectic invariant, but not the Lagrangian or the action.

Because the gauge fields in the supergravity action \eqref{S4dSG} are coupled to scalars, the duality transformation on the former requires also a transformation on the latter. This implies the existence of a flat symplectic bundle over the scalar manifold $\mathcal{M}_V$, that thus turns out to be a Special K\"ahler manifold \cite{Andrianopoli:2006ub}. Therefore, if we want the symplectic transformation to be a symmetry of the whole set of equations of motions, we need to restrict the transformations to those corresponding to isometries of the scalar manifold $\mathcal{M}_V$. In the $STU$ model the scalar manifold is $[SU(1,1)/U(1)]^3$, and in more general theories it is typically (thought not always) a coset space $\mathcal{M}_V=G/H$. The duality group is then the global symmetry group $G\subset Sp(2n_V+2,\mathbb{R})$.\footnote{In the microscopic string theory description of black holes as D-branes bound states, because of charge quantization the duality group is actually embedded in $Sp(2n_V+2,{\mathbb Z})$.}

We also recall the transformation rules for the scalar fields (see e.g.~\cite{deWit:1992wf}). Because of the special geometry structure of the scalar manifold, the scalar fields of the vector multiplets are equivalently parameterized by symplective projective coordinates \eqref{proj-coord}. Duality transformations act linearly on the symplectic section \eqref{Omega}, therefore
\begin{eqnarray}
\tilde X{}^\Lambda&=&U^\Lambda_{\ \Sigma}\,X^\Sigma + Z^{\Lambda\Sigma}\,
F_\Sigma,\nonumber\\
\tilde F{}_\Lambda&=&V_\Lambda{}^\Sigma\,F_\Sigma + W_{\Lambda\Sigma }\,X^\Sigma .
\label{xxx}
\end{eqnarray}
Owing to the symplectic conditions (\ref{spc}) and with one more assumption on $\tilde F_{\Lambda \Sigma}\equiv \partial_\Lambda \tilde F_\Sigma$ (it has to be symmetric), the quantities $\tilde F_\Lambda$ can be written as the derivative of a new function $\tilde F(\tilde X)$ with respect to the new coordinate $\tilde
X^\Lambda$:
\begin{equation}
\tilde F(\tilde X)\
=\ {\textstyle{1\over2}} \big(U^{\rm T}W\big)_{\Lambda\Sigma}X^\Lambda X^\Sigma
+{\textstyle{1\over2}}
\big(U^{\rm T}V + W^{\rm T}  Z\big)_\Lambda{}^\Sigma X^\Lambda F_\Sigma
+{\textstyle{1\over2}} \big(Z^{\rm T}V\big){}^{\Lambda \Sigma}F_\Lambda F_\Sigma \,,
\label{Ftilde}
\end{equation}
where we made use of the homogeneity of $F$. Lagrangians  parameterized by $F(X)$ and $\tilde F(\tilde X)$ represent equivalent theories, at least within the Abelian sector. 

Finally, for a given charged black hole solution, the duality transformations act on the electric and magnetic charge vectors in an analogous way:
\begin{eqnarray}
\tilde p{}^\Lambda&=&U^\Lambda_{\ \Sigma}\,p^\Sigma + Z^{\Lambda\Sigma}\,
q_\Sigma,\nonumber\\
\tilde q_\Lambda&=&V_\Lambda{}^\Sigma\,q_\Sigma + W_{\Lambda\Sigma}\,p^\Sigma .
\label{xxxcharge}
\end{eqnarray}
Black holes with charges $q_\Lambda,p^\Lambda$ and $\tilde q_\Lambda,\tilde p^\Lambda$ belong to equivalent solutions, just obtained in a different symplectic duality frame.
This is also true for the corresponding entropies and temperatures ${\cal S}(p^\Lambda,q_\Lambda), {\cal T}(p^\Lambda,q_\Lambda)$ and $\tilde{\cal S}(\tilde p^\Lambda, \tilde q_\Lambda), \tilde{\cal T}(\tilde p^\Lambda, \tilde q_\Lambda)$.

\subsubsection*{Symplectic transformations in the $STU$ model}

Let us consider the action of a duality transformation, which is also a symplectic transformation of $Sp(2n_V+2,{\mathbb R})$, on entropy and temperature. Here we focus on the $STU$ model with $n_V=3$. The symplectic transformations do not change the solution, i.e., they must keep entropy and temperature invariant, just being expressed in a different symplectic duality frame. Consider as an example the following transformation
\begin{equation}
\tilde F_\Lambda=-X^\Lambda\, , \quad \tilde X^\Lambda=F_\Lambda\, ,
\end{equation}
which also exchanges all four electric and magnetic charges with each other.
As one can easily see, this transformation is inverting all three scalar fields, i.e. $\tilde S=1/S$, $\tilde T=1/T$, $\tilde U=1/U$,
and hence also the overall volume of the torus $T^6$ gets inverted: $\tilde{\mathcal{V}}=1/\mathcal{V}$.
For the black hole solution with four non-vanishing charges $q_0$ and $p^1,p^2,p^3$, the four transformed non-vanishing charges are $\tilde p^0$ and $\tilde q_1,\tilde q_2,\tilde q_3$.
Expressed in terms of the transformed charges, the volume $\tilde{\mathcal{V}}$ has the form
\begin{equation}
\tilde{\mathcal{V}}={(\tilde p^0)^{3/2}\over \sqrt{\tilde q_1\tilde q_2\tilde q_3}}\, .
\end{equation}
The entropy in the extremal case then simply looks in the dual symplectic frame like
\begin{equation}
\tilde {\cal S}=2\pi \sqrt{\tilde p^0 \tilde q_1 \tilde q_2 \tilde q_3}\, .
\end{equation}
Comparing $\tilde{\mathcal{V}}$ with $\tilde{\cal S}$ one immediately sees that the previous relations (\ref{Vextremalq}) and (\ref{Vextremalp}) between entropy and volume still hold, now with the relabelled charges, namely now for fixed
magnetic charge $\tilde p^0$ or fixed electric charges $\tilde q_i$, respectively:
\begin{equation}\boxed{
\tilde{\mathcal{V}} = 2\pi{(\tilde p^0)^2\over \tilde {\cal S}}= \frac{1}{(2\pi)^3}\frac{\tilde {{\cal S}}^3} {(\tilde q_1 \tilde q_2 \tilde q_3)^2}\, .}
\end{equation}
For the non-extremal case, we can play a similar trick: to switch to the dual symplectic frame, one has to substitute in the expressions for $a_0,b_0,a^i,b^i$ (see eqs.(\ref{a0b0}) and (\ref{a1b1})) the relevant electric and magnetic charges by the dual symplectic counter parts. Then, the relations (\ref{Vnonextremal}) and (\ref{Vnonextremalp}) between entropy, temperature and volume still hold. Therefore, all previous results about the large and small entropy and temperature limits also hold in the dual symplectic frame.

\bibliographystyle{JHEP}
\bibliography{papers.bib}

\providecommand{\href}[2]{#2}\begingroup\raggedright\begin{thebibliography}{10}

\bibitem{Vafa:2005ui}
C.~Vafa, \emph{{The String landscape and the swampland}},
  \href{https://arxiv.org/abs/hep-th/0509212}{{\tt hep-th/0509212}}.

\bibitem{Ooguri:2006in}
H.~Ooguri and C.~Vafa, \emph{{On the Geometry of the String Landscape and the
  Swampland}},
  \href{http://dx.doi.org/10.1016/j.nuclphysb.2006.10.033}{\emph{Nucl. Phys. B}
  {\bf 766} (2007) 21--33}, [\href{https://arxiv.org/abs/hep-th/0605264}{{\tt
  hep-th/0605264}}].

\bibitem{Palti:2019pca}
E.~Palti, \emph{{The Swampland: Introduction and Review}},
  \href{http://dx.doi.org/10.1002/prop.201900037}{\emph{Fortsch. Phys.} {\bf
  67} (2019) 1900037}, [\href{https://arxiv.org/abs/1903.06239}{{\tt
  1903.06239}}].

\bibitem{Bonnefoy:2019nzv}
Q.~Bonnefoy, L.~Ciambelli, D.~L\"ust and S.~L\"ust, \emph{{Infinite Black Hole
  Entropies at Infinite Distances and Tower of States}},
  \href{http://dx.doi.org/10.1016/j.nuclphysb.2020.115112}{\emph{Nucl. Phys. B}
  {\bf 958} (2020) 115112}, [\href{https://arxiv.org/abs/1912.07453}{{\tt
  1912.07453}}].

\bibitem{DeBiasio:2020xkv}
D.~De~Biasio and D.~L{\"u}st, \emph{{Geometric Flow Equations for
  Schwarzschild-AdS Space-Time and Hawking-Page Phase Transition}},
  \href{http://dx.doi.org/10.1002/prop.202000053}{\emph{Fortsch. Phys.} {\bf
  68} (2020) 2000053}, [\href{https://arxiv.org/abs/2006.03076}{{\tt
  2006.03076}}].

\bibitem{Luben:2020wix}
M.~L{\"u}ben, D.~L{\"u}st and A.~R. Metidieri, \emph{{The Black Hole Entropy
  Distance Conjecture and Black Hole Evaporation}},
  \href{http://dx.doi.org/10.1002/prop.202000130}{\emph{Fortsch. Phys.} {\bf
  69} (2021) 2000130}, [\href{https://arxiv.org/abs/2011.12331}{{\tt
  2011.12331}}].

\bibitem{Hamada:2021yxy}
Y.~Hamada, M.~Montero, C.~Vafa and I.~Valenzuela, \emph{{Finiteness and the
  Swampland}},  \href{https://arxiv.org/abs/2111.00015}{{\tt 2111.00015}}.

\bibitem{Holzhey:1991bx}
C.~F.~E. Holzhey and F.~Wilczek, \emph{{Black holes as elementary particles}},
  \href{http://dx.doi.org/10.1016/0550-3213(92)90254-9}{\emph{Nucl. Phys. B}
  {\bf 380} (1992) 447--477}, [\href{https://arxiv.org/abs/hep-th/9202014}{{\tt
  hep-th/9202014}}].

\bibitem{Preskill:1991tb}
J.~Preskill, P.~Schwarz, A.~D. Shapere, S.~Trivedi and F.~Wilczek,
  \emph{{Limitations on the statistical description of black holes}},
  \href{http://dx.doi.org/10.1142/S0217732391002773}{\emph{Mod. Phys. Lett. A}
  {\bf 6} (1991) 2353--2362}.

\bibitem{Maldacena:1998uz}
J.~M. Maldacena, J.~Michelson and A.~Strominger, \emph{{Anti-de Sitter
  fragmentation}},
  \href{http://dx.doi.org/10.1088/1126-6708/1999/02/011}{\emph{JHEP} {\bf 02}
  (1999) 011}, [\href{https://arxiv.org/abs/hep-th/9812073}{{\tt
  hep-th/9812073}}].

\bibitem{Page:2000dk}
D.~N. Page, \emph{{Thermodynamics of near extreme black holes}},  9, 2000.
\newblock \href{https://arxiv.org/abs/hep-th/0012020}{{\tt hep-th/0012020}}.

\bibitem{Heydeman:2020hhw}
M.~Heydeman, L.~V. Iliesiu, G.~J. Turiaci and W.~Zhao, \emph{{The statistical
  mechanics of near-BPS black holes}},
  \href{http://dx.doi.org/10.1088/1751-8121/ac3be9}{\emph{J. Phys. A} {\bf 55}
  (2022) 014004}, [\href{https://arxiv.org/abs/2011.01953}{{\tt 2011.01953}}].

\bibitem{Agrawal:2020xek}
P.~Agrawal, S.~Gukov, G.~Obied and C.~Vafa, \emph{{Topological Gravity as the
  Early Phase of Our Universe}},  \href{https://arxiv.org/abs/2009.10077}{{\tt
  2009.10077}}.

\bibitem{Draper:2019utz}
P.~Draper and S.~Farkas, \emph{{Transplanckian Censorship and the Local
  Swampland Distance Conjecture}},
  \href{http://dx.doi.org/10.1007/JHEP01(2020)133}{\emph{JHEP} {\bf 01} (2020)
  133}, [\href{https://arxiv.org/abs/1910.04804}{{\tt 1910.04804}}].

\bibitem{Lanza:2021udy}
S.~Lanza, F.~Marchesano, L.~Martucci and I.~Valenzuela, \emph{{The EFT stringy
  viewpoint on large distances}},
  \href{http://dx.doi.org/10.1007/JHEP09(2021)197}{\emph{JHEP} {\bf 09} (2021)
  197}, [\href{https://arxiv.org/abs/2104.05726}{{\tt 2104.05726}}].

\bibitem{PhysRev.160.1113}
B.~S. DeWitt, \emph{Quantum theory of gravity. i. the canonical theory},
  \href{http://dx.doi.org/10.1103/PhysRev.160.1113}{\emph{Phys. Rev.} {\bf 160}
  (Aug, 1967) 1113--1148}.

\bibitem{Li:2021gbg}
Y.~Li, \emph{{Black holes and the swampland: the deep throat revelations}},
  \href{http://dx.doi.org/10.1007/JHEP06(2021)065}{\emph{JHEP} {\bf 06} (2021)
  065}, [\href{https://arxiv.org/abs/2102.04480}{{\tt 2102.04480}}].

\bibitem{Li:2021utg}
Y.~Li, \emph{{An Alliance in the Tripartite Conflict over Moduli Space}},
  \href{https://arxiv.org/abs/2112.03281}{{\tt 2112.03281}}.

\bibitem{Elander:2020rgv}
D.~Elander, A.~F. Faedo, D.~Mateos and J.~G. Subils, \emph{{Phase transitions
  in a three-dimensional analogue of Klebanov-Strassler}},
  \href{http://dx.doi.org/10.1007/JHEP06(2020)131}{\emph{JHEP} {\bf 06} (2020)
  131}, [\href{https://arxiv.org/abs/2002.08279}{{\tt 2002.08279}}].

\bibitem{Atick:1988si}
J.~J. Atick and E.~Witten, \emph{{The Hagedorn Transition and the Number of
  Degrees of Freedom of String Theory}},
  \href{http://dx.doi.org/10.1016/0550-3213(88)90151-4}{\emph{Nucl. Phys. B}
  {\bf 310} (1988) 291--334}.

\bibitem{Kounnas:1989dk}
C.~Kounnas and B.~Rostand, \emph{{Coordinate Dependent Compactifications and
  Discrete Symmetries}},
  \href{http://dx.doi.org/10.1016/0550-3213(90)90543-M}{\emph{Nucl. Phys. B}
  {\bf 341} (1990) 641--665}.

\bibitem{Angelantonj:2008fz}
C.~Angelantonj, C.~Kounnas, H.~Partouche and N.~Toumbas, \emph{{Resolution of
  Hagedorn singularity in superstrings with gravito-magnetic fluxes}},
  \href{http://dx.doi.org/10.1016/j.nuclphysb.2008.10.010}{\emph{Nucl. Phys. B}
  {\bf 809} (2009) 291--307}, [\href{https://arxiv.org/abs/0808.1357}{{\tt
  0808.1357}}].

\bibitem{Saraikin:2007jc}
K.~Saraikin and C.~Vafa, \emph{{Non-supersymmetric black holes and topological
  strings}},
  \href{http://dx.doi.org/10.1088/0264-9381/25/9/095007}{\emph{Class. Quant.
  Grav.} {\bf 25} (2008) 095007},
  [\href{https://arxiv.org/abs/hep-th/0703214}{{\tt hep-th/0703214}}].

\bibitem{Ivashchuk:1999jd}
V.~D. Ivashchuk and V.~N. Melnikov, \emph{{P-Brane black holes for general
  intersections}}, {\emph{Grav. Cosmol.} {\bf 5} (1999) 313--318},
  [\href{https://arxiv.org/abs/gr-qc/0002085}{{\tt gr-qc/0002085}}].

\bibitem{Abishev:2015pqa}
M.~E. Abishev, K.~A. Boshkayev, V.~D. Dzhunushaliev and V.~D. Ivashchuk,
  \emph{{Dilatonic dyon black hole solutions}},
  \href{http://dx.doi.org/10.1088/0264-9381/32/16/165010}{\emph{Class. Quant.
  Grav.} {\bf 32} (2015) 165010}, [\href{https://arxiv.org/abs/1504.07657}{{\tt
  1504.07657}}].

\bibitem{Loges:2019jzs}
G.~J. Loges, T.~Noumi and G.~Shiu, \emph{{Thermodynamics of 4D Dilatonic Black
  Holes and the Weak Gravity Conjecture}},
  \href{http://dx.doi.org/10.1103/PhysRevD.102.046010}{\emph{Phys. Rev. D} {\bf
  102} (2020) 046010}, [\href{https://arxiv.org/abs/1909.01352}{{\tt
  1909.01352}}].

\bibitem{Gibbons:1987ps}
G.~W. Gibbons and K.-i. Maeda, \emph{{Black Holes and Membranes in Higher
  Dimensional Theories with Dilaton Fields}},
  \href{http://dx.doi.org/10.1016/0550-3213(88)90006-5}{\emph{Nucl. Phys. B}
  {\bf 298} (1988) 741--775}.

\bibitem{Garfinkle:1990qj}
D.~Garfinkle, G.~T. Horowitz and A.~Strominger, \emph{{Charged black holes in
  string theory}},
  \href{http://dx.doi.org/10.1103/PhysRevD.43.3140}{\emph{Phys. Rev. D} {\bf
  43} (1991) 3140}.

\bibitem{Cheng:1993wp}
G.-J. Cheng, R.-R. Hsu and W.-F. Lin, \emph{{Dyonic black holes in string
  theory}}, \href{http://dx.doi.org/10.1063/1.530817}{\emph{J. Math. Phys.}
  {\bf 35} (1994) 4839--4847},
  [\href{https://arxiv.org/abs/hep-th/9302065}{{\tt hep-th/9302065}}].

\bibitem{Heidenreich:2015nta}
B.~Heidenreich, M.~Reece and T.~Rudelius, \emph{{Sharpening the Weak Gravity
  Conjecture with Dimensional Reduction}},
  \href{http://dx.doi.org/10.1007/JHEP02(2016)140}{\emph{JHEP} {\bf 02} (2016)
  140}, [\href{https://arxiv.org/abs/1509.06374}{{\tt 1509.06374}}].

\bibitem{Lee:2019wij}
S.-J. Lee, W.~Lerche and T.~Weigand, \emph{{Emergent Strings from Infinite
  Distance Limits}},  \href{https://arxiv.org/abs/1910.01135}{{\tt
  1910.01135}}.

\bibitem{Andrianopoli:2006ub}
L.~Andrianopoli, R.~D'Auria, S.~Ferrara and M.~Trigiante, \emph{{Extremal black
  holes in supergravity}}, {\emph{Lect. Notes Phys.} {\bf 737} (2008)
  661--727}, [\href{https://arxiv.org/abs/hep-th/0611345}{{\tt
  hep-th/0611345}}].

\bibitem{Behrndt:1996jn}
K.~Behrndt, G.~Lopes~Cardoso, B.~de~Wit, R.~Kallosh, D.~L{\"u}st and
  T.~Mohaupt, \emph{{Classical and quantum N=2 supersymmetric black holes}},
  \href{http://dx.doi.org/10.1016/S0550-3213(97)00028-X}{\emph{Nucl. Phys. B}
  {\bf 488} (1997) 236--260}, [\href{https://arxiv.org/abs/hep-th/9610105}{{\tt
  hep-th/9610105}}].

\bibitem{Greene:1996cy}
B.~R. Greene, \emph{{String theory on Calabi-Yau manifolds}},  in
  \emph{{Theoretical Advanced Study Institute in Elementary Particle Physics
  (TASI 96): Fields, Strings, and Duality}}, 6, 1996.
\newblock \href{https://arxiv.org/abs/hep-th/9702155}{{\tt hep-th/9702155}}.

\bibitem{Bohm:1999uk}
R.~Bohm, H.~Gunther, C.~Herrmann and J.~Louis, \emph{{Compactification of type
  IIB string theory on Calabi-Yau threefolds}},
  \href{http://dx.doi.org/10.1016/S0550-3213(99)00796-8}{\emph{Nucl. Phys. B}
  {\bf 569} (2000) 229--246}, [\href{https://arxiv.org/abs/hep-th/9908007}{{\tt
  hep-th/9908007}}].

\bibitem{Denef:1999idt}
F.~Denef, \emph{{Low energy physics from type IIB string theory}}.
\newblock PhD thesis, Leuven U., 1999.

\bibitem{Shmakova:1996nz}
M.~Shmakova, \emph{{Calabi-Yau black holes}},
  \href{http://dx.doi.org/10.1103/PhysRevD.56.R540}{\emph{Phys. Rev. D} {\bf
  56} (1997) 540--544}, [\href{https://arxiv.org/abs/hep-th/9612076}{{\tt
  hep-th/9612076}}].

\bibitem{Cvetic:1995uj}
M.~Cvetic and D.~Youm, \emph{{Dyonic BPS saturated black holes of heterotic
  string on a six torus}},
  \href{http://dx.doi.org/10.1103/PhysRevD.53.R584}{\emph{Phys. Rev. D} {\bf
  53} (1996) 584--588}, [\href{https://arxiv.org/abs/hep-th/9507090}{{\tt
  hep-th/9507090}}].

\bibitem{Cvetic:1995bj}
M.~Cvetic and A.~A. Tseytlin, \emph{{Solitonic strings and BPS saturated dyonic
  black holes}}, \href{http://dx.doi.org/10.1103/PhysRevD.53.5619}{\emph{Phys.
  Rev. D} {\bf 53} (1996) 5619--5633},
  [\href{https://arxiv.org/abs/hep-th/9512031}{{\tt hep-th/9512031}}].

\bibitem{Breitenlohner:1987dg}
P.~Breitenlohner, D.~Maison and G.~W. Gibbons, \emph{{Four-Dimensional Black
  Holes from Kaluza-Klein Theories}},
  \href{http://dx.doi.org/10.1007/BF01217967}{\emph{Commun. Math. Phys.} {\bf
  120} (1988) 295}.

\bibitem{Galli:2011fq}
P.~Galli, T.~Ortin, J.~Perz and C.~S. Shahbazi, \emph{{Non-extremal black holes
  of N=2, d=4 supergravity}},
  \href{http://dx.doi.org/10.1007/JHEP07(2011)041}{\emph{JHEP} {\bf 07} (2011)
  041}, [\href{https://arxiv.org/abs/1105.3311}{{\tt 1105.3311}}].

\bibitem{Bellucci:2008sv}
S.~Bellucci, S.~Ferrara, A.~Marrani and A.~Yeranyan, \emph{{stu Black Holes
  Unveiled}}, \href{http://dx.doi.org/10.3390/e10040507}{\emph{Entropy} {\bf
  10} (2008) 507}, [\href{https://arxiv.org/abs/0807.3503}{{\tt 0807.3503}}].

\bibitem{Ferrara:1996dd}
S.~Ferrara and R.~Kallosh, \emph{{Supersymmetry and attractors}},
  \href{http://dx.doi.org/10.1103/PhysRevD.54.1514}{\emph{Phys. Rev. D} {\bf
  54} (1996) 1514--1524}, [\href{https://arxiv.org/abs/hep-th/9602136}{{\tt
  hep-th/9602136}}].

\bibitem{Ceresole:2007rq}
A.~Ceresole, S.~Ferrara and A.~Marrani, \emph{{4d/5d Correspondence for the
  Black Hole Potential and its Critical Points}},
  \href{http://dx.doi.org/10.1088/0264-9381/24/22/023}{\emph{Class. Quant.
  Grav.} {\bf 24} (2007) 5651--5666},
  [\href{https://arxiv.org/abs/0707.0964}{{\tt 0707.0964}}].

\bibitem{Gibbons:1985ac}
G.~W. Gibbons and D.~L. Wiltshire, \emph{{Black Holes in Kaluza-Klein Theory}},
  \href{http://dx.doi.org/10.1016/S0003-4916(86)80012-4}{\emph{Annals Phys.}
  {\bf 167} (1986) 201--223}.

\bibitem{Horowitz:2011cq}
G.~T. Horowitz and T.~Wiseman, \emph{{General black holes in
  Kaluza\textendash{}Klein theory}}, pp.~69--98.
\newblock 2012.
\newblock \href{https://arxiv.org/abs/1107.5563}{{\tt 1107.5563}}.

\bibitem{Duff:1999gh}
M.~J. Duff and J.~T. Liu, \emph{{Anti-de Sitter black holes in gauged N = 8
  supergravity}},
  \href{http://dx.doi.org/10.1016/S0550-3213(99)00299-0}{\emph{Nucl. Phys. B}
  {\bf 554} (1999) 237--253}, [\href{https://arxiv.org/abs/hep-th/9901149}{{\tt
  hep-th/9901149}}].

\bibitem{Ceresole:2009jc}
A.~Ceresole, S.~Ferrara, A.~Gnecchi and A.~Marrani, \emph{{More on N=8
  Attractors}}, \href{http://dx.doi.org/10.1103/PhysRevD.80.045020}{\emph{Phys.
  Rev. D} {\bf 80} (2009) 045020}, [\href{https://arxiv.org/abs/0904.4506}{{\tt
  0904.4506}}].

\bibitem{Ferrara:2006em}
S.~Ferrara and R.~Kallosh, \emph{{On N=8 attractors}},
  \href{http://dx.doi.org/10.1103/PhysRevD.73.125005}{\emph{Phys. Rev. D} {\bf
  73} (2006) 125005}, [\href{https://arxiv.org/abs/hep-th/0603247}{{\tt
  hep-th/0603247}}].

\bibitem{Bena:2009ev}
I.~Bena, G.~Dall'Agata, S.~Giusto, C.~Ruef and N.~P. Warner, \emph{{Non-BPS
  Black Rings and Black Holes in Taub-NUT}},
  \href{http://dx.doi.org/10.1088/1126-6708/2009/06/015}{\emph{JHEP} {\bf 06}
  (2009) 015}, [\href{https://arxiv.org/abs/0902.4526}{{\tt 0902.4526}}].

\bibitem{Ceresole:2010nm}
A.~Ceresole, S.~Ferrara and A.~Marrani, \emph{{Small N=2 Extremal Black Holes
  in Special Geometry}},
  \href{http://dx.doi.org/10.1016/j.physletb.2010.08.053}{\emph{Phys. Lett. B}
  {\bf 693} (2010) 366--372}, [\href{https://arxiv.org/abs/1006.2007}{{\tt
  1006.2007}}].

\bibitem{Lust:2019zwm}
D.~L{\"u}st, E.~Palti and C.~Vafa, \emph{{AdS and the Swampland}},
  \href{http://dx.doi.org/10.1016/j.physletb.2019.134867}{\emph{Phys. Lett. B}
  {\bf 797} (2019) 134867}, [\href{https://arxiv.org/abs/1906.05225}{{\tt
  1906.05225}}].

\bibitem{Tsimpis:2012tu}
D.~Tsimpis, \emph{{Supersymmetric AdS vacua and separation of scales}},
  \href{http://dx.doi.org/10.1007/JHEP08(2012)142}{\emph{JHEP} {\bf 08} (2012)
  142}, [\href{https://arxiv.org/abs/1206.5900}{{\tt 1206.5900}}].

\bibitem{Gautason:2015tig}
F.~F. Gautason, M.~Schillo, T.~Van~Riet and M.~Williams, \emph{{Remarks on
  scale separation in flux vacua}},
  \href{http://dx.doi.org/10.1007/JHEP03(2016)061}{\emph{JHEP} {\bf 03} (2016)
  061}, [\href{https://arxiv.org/abs/1512.00457}{{\tt 1512.00457}}].

\bibitem{Lust:2020npd}
D.~L{\"u}st and D.~Tsimpis, \emph{{AdS$_{2}$ type-IIA solutions and scale
  separation}}, \href{http://dx.doi.org/10.1007/JHEP07(2020)060}{\emph{JHEP}
  {\bf 07} (2020) 060}, [\href{https://arxiv.org/abs/2004.07582}{{\tt
  2004.07582}}].

\bibitem{Behrndt:1997ny}
K.~Behrndt, D.~L{\"u}st and W.~A. Sabra, \emph{{Stationary solutions of N=2
  supergravity}},
  \href{http://dx.doi.org/10.1016/S0550-3213(97)00633-0}{\emph{Nucl. Phys. B}
  {\bf 510} (1998) 264--288}, [\href{https://arxiv.org/abs/hep-th/9705169}{{\tt
  hep-th/9705169}}].

\bibitem{Bena:2007kg}
I.~Bena and N.~P. Warner, \emph{{Black holes, black rings and their
  microstates}},
  \href{http://dx.doi.org/10.1007/978-3-540-79523-0_1}{\emph{Lect. Notes Phys.}
  {\bf 755} (2008) 1--92}, [\href{https://arxiv.org/abs/hep-th/0701216}{{\tt
  hep-th/0701216}}].

\bibitem{Cribiori:2021gbf}
N.~Cribiori, D.~Lust and M.~Scalisi, \emph{{The gravitino and the swampland}},
  \href{http://dx.doi.org/10.1007/JHEP06(2021)071}{\emph{JHEP} {\bf 06} (2021)
  071}, [\href{https://arxiv.org/abs/2104.08288}{{\tt 2104.08288}}].

\bibitem{Castellano:2021yye}
A.~Castellano, A.~Font, A.~Herraez and L.~E. Ib\'a\~nez, \emph{{A gravitino
  distance conjecture}},
  \href{http://dx.doi.org/10.1007/JHEP08(2021)092}{\emph{JHEP} {\bf 08} (2021)
  092}, [\href{https://arxiv.org/abs/2104.10181}{{\tt 2104.10181}}].

\bibitem{Andrianopoli:1996cm}
L.~Andrianopoli, M.~Bertolini, A.~Ceresole, R.~D'Auria, S.~Ferrara, P.~Fre
  et~al., \emph{{N=2 supergravity and N=2 superYang-Mills theory on general
  scalar manifolds: Symplectic covariance, gaugings and the momentum map}},
  \href{http://dx.doi.org/10.1016/S0393-0440(97)00002-8}{\emph{J. Geom. Phys.}
  {\bf 23} (1997) 111--189}, [\href{https://arxiv.org/abs/hep-th/9605032}{{\tt
  hep-th/9605032}}].

\bibitem{Ferrara:1995ih}
S.~Ferrara, R.~Kallosh and A.~Strominger, \emph{{N=2 extremal black holes}},
  \href{http://dx.doi.org/10.1103/PhysRevD.52.R5412}{\emph{Phys. Rev. D} {\bf
  52} (1995) R5412--R5416}, [\href{https://arxiv.org/abs/hep-th/9508072}{{\tt
  hep-th/9508072}}].

\bibitem{Ferrara:1996um}
S.~Ferrara and R.~Kallosh, \emph{{Universality of supersymmetric attractors}},
  \href{http://dx.doi.org/10.1103/PhysRevD.54.1525}{\emph{Phys. Rev. D} {\bf
  54} (1996) 1525--1534}, [\href{https://arxiv.org/abs/hep-th/9603090}{{\tt
  hep-th/9603090}}].

\bibitem{Ceresole:1995jg}
A.~Ceresole, R.~D'Auria, S.~Ferrara and A.~Van~Proeyen, \emph{{Duality
  transformations in supersymmetric Yang-Mills theories coupled to
  supergravity}},
  \href{http://dx.doi.org/10.1016/0550-3213(95)00175-R}{\emph{Nucl. Phys. B}
  {\bf 444} (1995) 92--124}, [\href{https://arxiv.org/abs/hep-th/9502072}{{\tt
  hep-th/9502072}}].

\bibitem{deWit:1995dmj}
B.~de~Wit, V.~Kaplunovsky, J.~Louis and D.~L{\"u}st, \emph{{Perturbative
  couplings of vector multiplets in N=2 heterotic string vacua}},
  \href{http://dx.doi.org/10.1016/0550-3213(95)00291-Y}{\emph{Nucl. Phys. B}
  {\bf 451} (1995) 53--95}, [\href{https://arxiv.org/abs/hep-th/9504006}{{\tt
  hep-th/9504006}}].

\bibitem{deWit:1992wf}
B.~de~Wit, F.~Vanderseypen and A.~Van~Proeyen, \emph{{Symmetry structure of
  special geometries}},
  \href{http://dx.doi.org/10.1016/0550-3213(93)90413-J}{\emph{Nucl. Phys. B}
  {\bf 400} (1993) 463--524}, [\href{https://arxiv.org/abs/hep-th/9210068}{{\tt
  hep-th/9210068}}].

\end{thebibliography}\endgroup

\end{document}